\documentclass[11pt]{article}
\usepackage{mathrsfs}
\usepackage{amsmath}
\usepackage{amsfonts}
\usepackage{amssymb, amsmath, cite}
\usepackage{color}
\usepackage{graphicx}

\newtheorem{theorem}{Theorem}

\newcommand\be{\begin{equation}}
\newcommand\ee{\end{equation}}
\newcommand\ber{\begin{eqnarray}}
\newcommand\eer{\end{eqnarray}}
\newcommand\berr{\begin{eqnarray*}}
\newcommand\eerr{\end{eqnarray*}}
\newcommand\bea{\begin{eqnarray}}
\newcommand\eea{\end{eqnarray}}

\newcommand{\bfR}{{\Bbb R}}

\newcommand{\x}{{\bf x}}

\newcommand{\xx}{{\bf x}}\newcommand{\rr}{{\bf r}}

\newcommand{\dd}{\mbox{d}}\newcommand{\D}{\cal O}
\newcommand{\e}{\mbox{e}}
\newcommand{\pa}{\partial}
\newcommand{\vep}{\varepsilon}

\newcommand{\nn}{\nonumber}

\newcommand\lb{\label}
\newcommand\eq{\eqref}

\title{Exact Multicentered Static Point Charge Source Distributions\\ in the Born--Infeld Theory of Nonlinear Electromagnetism\\ and Presence of Electric and Magnetic Currents}

\author{
Yisong Yang\footnote{Email address: yisongyang@nyu.edu}\\Courant Institute of Mathematical Sciences\\ New York University\\New York, New York 10012, USA
}

\date{}

\begin{document}

\maketitle

\begin{abstract}
Exact multicentered solutions to the governing equations of the Born--Infeld nonlinear theory of electrodynamics  describing distributions of electric, magnetic, and
dyonic point charge sources are constructed explicitly for the first time. The method of construction may effectively be adapted to obtain such solutions in
generalized theories as well. As a consequence, the solutions unveil that, in order to achieve a balance between such multicentered electric or magnetic point charge
distribution in equilibrium, a static magnetic or electric current must be present, resulting in non-conservativeness of the induced electric or magnetic intensity field, respectively and universally. Moreover, it is also shown that the methods of construction and conclusions drawn in the multicentered situation
may be extended to obtain exact solutions and similar conclusions, explicitly, in the situation of continuously distributed charge source problems.
\medskip

{\flushleft {Keywords:} Born--Infeld nonlinear electrodynamics,  exact multicentered solutions, finite-energy condition, electric and magnetic currents, 
non-conservativeness of fields.
\medskip

{PACS numbers:} 02.30.Jr, 02.90.$+$p, 03.50.$-$z, 11.10.Lm
\medskip

{MSC numbers:} 35C05, 35Q60, 78A25
}
\end{abstract}

\section{Introduction}
\setcounter{equation}{0}

In a series of works \cite{Sch1,Sch2,Sch3} initiated in 1960's, Schwinger advanced a fundamental subject aimed at formulating an elementary particle model built over
the concept of  electrically and magnetically charged particles called dyons. Schwinger's main idea was to exploit the electric and magnetic symmetry, uniquely exhibited in four
spacetime dimensions
through the Hodge duality, realizing electromagnetic duality, in the Maxwell equations. Use $\bf E$ and $\bf B$ to denote the electric and magnetic fields.
Subject to some hypothetically distributed electric and magnetic charge densities $\rho_e$ and $\rho_m$ and electric and magnetic current densities ${\bf j}_e$ and ${\bf j}_m$,
respectively, Schwinger's covariant Maxwell equations are 
\be\lb{01}
\frac{\pa {\bf B}}{\pa t}+\nabla\times{\bf E}=-{\bf j}_m,\quad \nabla\cdot{\bf B}=\rho_m,\quad \frac{\pa{\bf E}}{\pa t}-\nabla\times{\bf B}=-{\bf j}_e,\quad \nabla\cdot{\bf E}=\rho_e.
\ee
See also \cite{Z,Jackson,Bla,Sin,Singleton,Mig,Milton}. In the classical Maxwell equations, magnetic current and charge densities are both
absent so that the first equation in \eq{01} with ${\bf j}_m={\bf0} $ is Faraday's law and the second equation in \eq{01} with $\rho_m=0$ the Gauss law, which are constraint equations resulting from the Bianchi identity. Such a system has a natural variational formalism based on the gauge-field-theoretical 
action density
\be\lb{x1.2}
{\cal L}=-\frac14F_{\mu\nu}F^{\mu\nu}-A_\mu j^\mu,\quad F_{\mu\nu}=\pa_\mu A_\nu-\pa_\nu A_\mu,\quad j^\mu =(\rho_e,{\bf j}_e),
\ee
where the electric source terms, $\rho_e$ and ${\bf j}_e$, are directly coupled with the electromagnetic gauge potential field $A_\mu$, but there is no
room for their magnetic counterparts, $\rho_m$ and ${\bf j}_m$. 
In Schwinger's extended system, however, these magnetic source terms are switched on or {\em imposed} to restore  electromagnetic symmetry or duality \cite{Sch1,Sch2,Sch3} so that the Bianchi identity is
violated, inevitably, and the system does not enjoy a similar variational or gauge-field-theoretical formalism. In fact, it is well known that
the gauge potential
field origin of the electric and magnetic fields in the classical Maxwell theory follows from the homogeneity condition
comprised of $\rho_m=0$ and ${\bf j}_m={\bf 0}$. 
 In \cite{Sch1,Sch2}, Schwinger considered the quantum mechanical motion of a dyonic point charge carrying electric charge $q_1$ and magnetic charge $g_1$ in a static
electromagnetic field generated by another dyonic point charge at rest at the origin of $\bfR^3$ carrying electric charge $q_2$ and magnetic charge $g_2$ such that the electric and
magnetic fields, $\bf E$ and $\bf B$,  it generates, are given by
\be\lb{02}
{\bf E}=\frac{q_2{\bf x}}{|{\bf x}|^3},\quad {\bf B}=\frac{g_2{\bf x}}{|{\bf x}|^3},\quad {\bf x}\in\bfR^3,
\ee
 and thereby derived the charge quantization condition
\be
q_1g_2-q_2 g_1=\frac{n\hbar}2,
\ee
where $n$ is an integer, which generalizes Dirac's charge quantization formula \cite{Dirac} based on describing the quantum mechanical motion of an electric point charge in a monopole field.
As a consequence of the point charge solution \eq{02}, it follows from \eq{01} that
both electric and magnetic currents must vanish
\be\lb{04}
{\bf j}_e={\bf 0},\quad {\bf j}_m={\bf 0},
\ee
to ensure consistency with $\nabla\times{\bf E}={\bf0}$ and $\nabla\times{\bf B}={\bf0}$, since $\bf E$ and $\bf B$ given in \eq{02} are both conservative. Due to the linearity of the generalized Maxwell equations \eq{01}, the same conclusion, \eq{04}, holds
for multicentered source situations. In other words, in this context, multicentered charges are accommodated, which necessarily switch off the currents, as in the singly-centered situation.

In the early 1930's, Born and Infeld \cite{B1,B2,B3,B4} formulated a nonlinear theory of electromagnetism aimed at overcoming the infinity problem associated with an electric point charge.
In contemporary theoretical physics, the Born--Infeld theory and its generalized forms are seen to arise in the studies on superstrings \cite{FT,Ts2,Tsey}  and branes \cite{CM,Gibbons,Ts1}, charged black holes \cite{AG1,AG2,K1,K2,Yang1,Yang2,Yang3},
and cosmology \cite{Jana,Kam,Nov,Yang1,Yang3}. See \cite{JHOR} for a review.  Due to the difficulties associated with the nonlinearity of the Born--Infeld type theories, the charged
source distributions in these and other studies have exclusively been based on singly-centered point charges. The main contribution of the present work is to obtain, for the first time,
multicentered static point charge distributions as exact and explicit solutions of the Born--Infeld equations, and their extensions. Besides being successful in producing finite-energy electric
point charge as achieved in the original Born--Infeld work \cite{B1,B2,B3,B4}, numerous important insights offered by this type of nonlinear theories of electrodynamics have been garnered
in development. These include a construction of charged black holes with removed or relegated singularities \cite{AG1,AG2,Yang1,Yang2,JHOR,Yang3}, a mechanism for the exclusion of monopoles \cite{Yang1,Yang4}, and a formalism for a theoretical interpretation of the equation of state of a prescribed k-essence cosmic fluid \cite{Yang1,Yang3,Yang4}. As a by-product,
the current study offers yet another interesting insight injected by the subject: Due to the nonlinear nature of the theory, the electric or magnetic intensity field generated from a multicentered electric or magnetic point charge source must be
non-conservative, characterized by a violation of ${\bf j}_m={\bf 0}$ or ${\bf j}_e={\bf0}$, in a sharp contrast against \eq{04}, valid for Schwinger's generalized Maxwell equations, \eq{01}.
In other words, we observe that the Born--Infeld type nonlinear theories of electrodynamics possess a tight internal coupling between electricity and magnetism in that the presence of a 
multicentered electric point charge source leads to the onset of a static magnetic current, and that of magnetic point charge source the onset of a static electric current, although such a 
phenomenon does not appear in the singly-centered point charge situations of the theories.

The rest of the presentation of this work is organized as follows. In Section \ref{sec2}, we briefly review the Born--Infeld theory and state our main results. In Section \ref{sec3}, we
construct the exact and explicit solutions to the Born--Infeld equations representing multicentered electric and magnetic point charge sources and demonstrate the non-conservativeness of
the electric and magnetic fields generated. In Section \ref{sec4}, we extend our consideration to multicentered dyonically charged solutions. In Sections \ref{sec5}--\ref{sec6}, we
show that our methods can be adapted to construct multicentered solutions in generalized Born--Infeld theories and that we can draw the same conclusions. In particular, we see that in either
the multicentered electric or magnetic point charge source situation, electric charge and magnetic current, or magnetic charge and electric current, must coexist in a {\em nonlinear} theory of electrodynamics of the Born--Infeld type. In Section 7, we extend our work to the situation of continuously distributed charge source problems and obtain exact solutions and a series of similar conclusions on such problems. In Section 8, we summarize the results.

\section{Born--Infeld theory of nonlinear electrodynamics and presence of electric and magnetic currents}\lb{sec2}
\setcounter{equation}{0}

We consider the Born--Infeld theory of nonlinear electrodynamics based on the free Lagrangian action density \cite{B4,Yang2,Yang3,Kr1}
\bea
&&{\cal L}=\frac1{\beta}\left(1-\sqrt{1-2\beta s}\right),\lb{1.1}\\
&&s=-\frac14F_{\mu\nu}F^{\mu\nu}+\frac{\kappa^2}{32}\left(F_{\mu\nu}\tilde{F}^{\mu\nu}\right)^2,\lb{x2.2}
\eea
where the electromagnetic tensor field $F^{\mu\nu}$ and its Hodge dual $\tilde{F}^{\mu\nu}=\frac12\epsilon^{\mu\nu\alpha\beta}F_{\alpha\beta}$ are
associated with the underlying electric field ${\bf E}=(E^i)$ and magnetic field ${\bf B}=(B^i)$ through the expressions
\be\lb{x2.3}
(F^{\mu\nu})=\left(\begin{array}{cccc}0&-E^1&-E^2&-E^3\\E^1&0&-B^3&B^2\\E^2&B^3&0&-B^1\\E^3&-B^2&B^1&0\end{array}\right),\quad
(\tilde{F}^{\mu\nu})=\left(\begin{array}{cccc}0&-B^1&-B^2&-B^3\\B^1&0&E^3&-E^2\\B^2&-E^3&0&E^1\\B^3&E^2&-E^1&0\end{array}\right),
\ee
the parameter $\beta>0$ relates to the usual Born parameter, and $\kappa>0$ is an electromagnetic coupling parameter that directly mixes the interaction of the fields ${\bf E}$ and ${\bf B}$.
In the presence of a source current $j^\mu$ as given in \eq{x1.2}, the action density \eq{1.1} is modified into
\be\lb{x2.4}
{\cal L}=\frac1{\beta}\left(1-\sqrt{1-2\beta s}\right)-A_\mu j^\mu,
\ee
so that varying the gauge field $A_\mu$ leads to its Euler--Lagrange equation
\be\lb{x2.5}
\pa_\mu P^{\mu\nu}=j^\nu,
\ee
where
\be\lb{x2.6}
P^{\mu\nu}=\frac1{\sqrt{1-2\beta s}}\left(F^{\mu\nu}-\frac{\kappa^2}4(F_{\alpha\beta}\tilde{F}^{\alpha\beta})\tilde{F}^{\mu\nu}\right)
\ee
gives rise to the associated electric displacement field $\bf D$ and magnetic intensity field $\bf H$ expressed in the matrix
\be\lb{x2.7}
(P^{\mu\nu})=\left(\begin{array}{cccc}0&-D^1&-D^2&-D^3\\D^1&0&-H^3&H^2\\D^2&H^3&0&-H^1\\D^3&-H^2&H^1&0\end{array}\right).
\ee
With \eq{x2.3}, we see that \eq{x2.2} becomes 
\be
s=\frac12({\bf E}^2-{\bf B}^2)+\frac{\kappa^2}2({\bf E}\cdot{\bf B})^2,\lb{1.2}
\ee
and that \eq{x2.6} leads to the associated nonlinear constitutive equations of the theory:
\bea
&&{\bf D}=\frac1{\sqrt{1-2\beta s}}({\bf E}+\kappa^2 [{\bf E}\cdot{\bf B}]{\bf B}),\lb{1.3}\\
&&{\bf H}=\frac1{\sqrt{1-2\beta s}}({\bf B}-\kappa^2 [{\bf E}\cdot{\bf B}]{\bf E}).\lb{1.4}
\eea
Note that these constitutive equations are the fundamental consequences of the variational principles or structure of the theory \cite{B1,B2,B3,B4} and independent of the source terms.

At this juncture, it will be relevant to recall the different roles played by the pairs ${\bf E},{\bf B}$ and ${\bf D},{\bf H}$, with respect to the
source terms. In fact, in terms of \eq{x2.5} and \eq{x2.7}, we have 
\be\lb{x2.11}
-\frac{\pa{\bf D}}{\pa t}+\nabla\times{\bf H}={\bf j}_e,\quad \nabla\cdot{\bf D}=\rho_e,
\ee
which are the Amper\'{e} law and Coulomb law, respectively, in the Born--Infeld setting. From the gauge field representation of $F_{\mu\nu}$ given in \eq{x1.2}, we see that there holds
the identity
\be\lb{x2.12}
\pa^\gamma F^{\mu\nu}+\pa^\mu F^{\nu\gamma}+\pa^\nu F^{\gamma\mu}=0, \quad\mbox{or}\quad \pa_\mu \tilde{F}^{\mu\nu}=0,
\ee
known as the Bianchi identity. Using \eq{x2.3}, we see that \eq{x2.12} gives rise to the equations
\be\lb{x2.13}
\frac{\pa{\bf B}}{\pa t}+\nabla\times{\bf E}={\bf 0},\quad\nabla\cdot{\bf B}=0,
\ee
which are the Faraday law and Gauss law, respectively, also in the Born--Infeld setting. The equations \eq{x2.11} and \eq{x2.13} are formally
identical to the full classical Maxwell equations, with a hidden nonlinear structure contained in \eq{1.3} and \eq{1.4}. In particular, as in the classical Maxwell electromagnetism,
we have observed that a gauge-field-theoretical formalism of the the Born--Infeld theory again prohibits the presence of magnetic source terms. 

As in Dirac \cite{Dirac} and Schwinger \cite{Sch1,Sch2,Sch3}, we may now subject the Born--Infeld equations involving ${\bf E},{\bf B}$
and ${\bf D},{\bf H}$ to applied electric and magnetic charge densities $\rho_e$ and $\rho_m$ and current
densities ${\bf j}_e$ and ${\bf j}_m$, respectively, in order to achieve electromagnetic duality, which necessarily turns off the gauge field formalism
when $\rho_m$ or ${\bf j}_m$ is nonvanishing, in view of \eq{x2.12} or \eq{x2.13}, as in \eq{01}, reviewed earlier. In such a context, the full equations of motion, which are the Born--Infeld theory version of the covariant Maxwell equation \eq{01}, are given by
\bea
&&\frac{\pa{\bf B}}{\pa t}+\nabla\times{\bf E}=-{\bf j}_m,\lb{1.5}\\
&&\frac{\pa{\bf D}}{\pa t}-\nabla\times{\bf H}=-{\bf j}_e,\lb{1.6}\\
&&\nabla\cdot{\bf D}=\rho_e,\lb{1.7}\\
&&\nabla\cdot{\bf B}=\rho_m.\lb{1.8}
\eea
In fact, these equations are formally exactly the same generalized Maxwell equations \cite{Sch1,Jackson,Singleton,Mig,Milton} when ${\bf E},{\bf B}$ and ${\bf D},{\bf H}$ are related through
the usual linear constitutive equations
\be
{\bf D}=\vep{\bf E},\quad {\bf B}=\mu{\bf H},
\ee
where the constants $\vep$ and $\mu$ are dielectrics and permeability coefficients, such that $c=\frac1{\sqrt{\vep\mu}}$ gives rise to the speed of light in the medium. In free space, this
quantity is normalized to unity, $c=1$. In particular, when $\vep$ and $\mu$ are both further normalized to unity, \eq{1.5}--\eq{1.8} return to \eq{01}.
We are interested in static solutions of \eq{1.5}--\eq{1.8} when $\rho_e$ and $\rho_m$ there are given as sums of the Dirac distributions realizing point charge sources.
Note that \eq{1.3} and \eq{1.4} may be interpreted as giving rise to a nonlinear medium with field-dependent dielectrics and permeability coefficients following the relation
\be\lb{1.9}
\left(\begin{array}{c}{\bf D}\\{\bf B}\end{array}\right)=\Sigma({\bf E},{\bf B})\left(\begin{array}{c}{\bf E}\\{\bf H}\end{array}\right),\quad
 \Sigma({\bf E},{\bf B})\equiv\left(\begin{array}{cc}\frac{1+\kappa^4({\bf E}\cdot{\bf B})^2}{\sqrt{1-2\beta s}}& \kappa^2({\bf E}\cdot{\bf B})\\ \kappa^2({\bf E}\cdot{\bf B})&\sqrt{1-2\beta s}\end{array}\right).
\ee
As a consequence, the condition $c=1$ is reflected by the property $\det(\Sigma({\bf E},{\bf B}))=1$. 

Below is our main theorem regarding the 
equations \eq{1.5}--\eq{1.8}.

\begin{theorem}\lb{th}
Consider the static solutions of the classical Born--Infeld equations \eq{1.5}--\eq{1.8} where the electric and magnetic fields $\bf E$ and $\bf B$ are related to the electric displacement
and magnetic intensity fields $\bf D$ and $\bf H$ through the constitutive equations \eq{1.3}--\eq{1.4} of the theory.
\begin{enumerate}
\item[(i)] When the electric charge density $\rho_e$ describes a multicentered electric point charge source, the electrostatic solution of the equations can be constructed explicitly and exactly
which is of finite energy, its free electric charge agrees with the total prescribed one, it is magnetically trivial characterized by ${\bf B}={\bf0}$ and ${\bf H}={\bf0}$, hence $\rho_m=0$
and ${\bf j}_e={\bf0}$, but its induced magnetic current density is turned on, ${\bf j}_m\neq{\bf0}$, which makes the electric field $\bf E$ non-conservative, $\nabla\times{\bf E}=-{\bf j}_m
\neq{\bf 0}$.

\item[(ii)] Likewise, when the magnetic charge density $\rho_m$ describes a multicentered magnetic point charge source,  similar conclusions concerning various fields about the magnetostatic
solution hold. In particular, the induced electric current density is now turned on, ${\bf j}_e\neq{\bf0}$, which renders the magnetic intensity field $\bf H$ non-conservative, 
$\nabla\times{\bf H}={\bf j}_e\neq{\bf 0}$.

\item[(iii)] In the full dyonic situation where $\rho_e$ and $\rho_m$ are both present to describe a multicentered dyonic point charge source, the solution of the equations can also be
constructed explicitly and exactly and is of a finite energy if and only if the electromagnetic coupling parameter $\kappa$ is nonzero. In this situation, the free electric and magnetic charges
are the same as the prescribed ones, $\bf E$ and $\bf H$ are both non-conservative, and ${\bf j}_e$ and ${\bf j}_m$ are both turned on.
\end{enumerate}
\end{theorem}

We emphasize that, unlike that in the covariant Maxwell equations \eq{01} of Schwinger \cite{Sch1,Sch2,Sch3}, the magnetic current ${\bf j}_m$, for example, is physically
{\em self-induced} by a multicentered electric charge source distribution, in order to maintain a balanced equilibrium state,
in nonlinear electrodynamics, rather than being manually imposed in linear electrodynamics.

Detailed properties of the solutions, including various field representations and charge and energy descriptions, will be made clear in the subsequent development of the subject.

\section{Electric and magnetic point charges}\lb{sec3}
\setcounter{equation}{0}

First consider the electrostatic case where
\be\lb{2.1}
\rho_e({\bf x})=\sum_{i=1}^n q_i \delta({\bf x}-{\bf x}_i)= \sum_{i=1}^n q_i \delta({\bf r}_i),\, {\bf x},{\bf x}_i\in\bfR^3,\, \rr_i=\xx-\xx_i,\, q_i\in\bfR\setminus\{0\},\, i=1,\dots,n.
\ee
It is consistent to set $\rho_m=0$ and ${\bf B}={\bf 0}$ in \eq{1.8}. Thus we have ${\bf H}={\bf0}$ by \eq{1.4}  and ${\bf j}_e={\bf 0}$ in \eq{1.6}. Inserting \eq{2.1}
into \eq{1.7}, we obtain
\be\lb{2.2}
{\bf D}=\sum_{i=1}^n \frac{q_i\rr_i}{4\pi|\rr_i|^3}=-\nabla U_e,\quad U_e({\bf x})=\sum_{i=1}^n \frac{q_i}{4\pi|\rr_i|},
\ee
where the factor $4\pi$ is inserted throughout to normalize the subsequent charge calculations, in contrast to \eq{02}.
On the other hand, squaring \eq{1.3}, we have the solution
\be
{\bf E}^2=\frac{{\bf D}^2}{1+\beta {\bf D}^2}.\lb{2.3}
\ee
Thus, inserting \eq{2.3} back into \eq{1.3} with $s=\frac12 {\bf E}^2$, we obtain the following explicit expression of the electric field ${\bf E}$:
\bea
{\bf E}&=&\frac{{\bf D}}{\sqrt{1+\beta{\bf D}^2}} \nn\\
&=&\sum_{i=1}^n \frac{q_i\rr_i}{4\pi|\rr_i|^3\sqrt{1+\beta \left(\sum_{j=1}^n \frac{q_j\rr_j}{4\pi|\rr_j|^3}\right)^2}}.\lb{2.4}
\eea
Hence, we have
\be\lb{2.5}
{\bf E}({\bf x})=\frac{q_i \rr_i}{4\pi |\rr_i|\sqrt{|\rr_i|^4+{\beta}\left(\frac{q_i}{4\pi}\right)^2}},\quad |\rr_i|\ll1,\quad i=1,\dots,n; \quad {\bf E}({\bf x})=\sum_{i=1}^n \frac{q_i\rr_i}{4\pi|\rr_i|^3},\quad
|{\bf x}|\gg1.
\ee
The free electric charge density induced from the electric field is given by
\be\lb{2.6}
\rho^{\mbox{\tiny free}}_e=\nabla\cdot{\bf E},
\ee
which is analogous to \eq{1.7}. From \eq{2.6} along with the divergence theorem and \eq{2.4}--\eq{2.5}, we find the total free electric charge $Q_{\mbox{\tiny free}}$ to be
\bea\lb{2.7}
Q_{\mbox{\tiny free}}&=&\int_{\bfR^3}\rho^{\mbox{\tiny free}}_e\,\dd{\bf x}\nn\\
&=&\lim_{R\to\infty}\sum_{i=1}^n \int_{|\rr_i|=R}\frac{q_i\rr_i\cdot \,\dd{\bf S}}{4\pi|\rr_i|^3\sqrt{1+\beta \left(\sum_{j=1}^n \frac{q_j\rr_j}{4\pi|\rr_j|^3}\right)^2}}\nn\\
&=&\sum_{i=1}^n q_i=\int_{\bfR^3}\rho_e\,\dd{\bf x}=Q.
\eea
That is, the {\em induced} total free electric charge agrees with the total electric charge given as a collection of prescribed multicentered electric point charges. 
Moreover, the electrostatic Hamiltonian energy density
of the model consisting of \eq{1.1} and \eq{1.2} reads
\bea\lb{2.8}
{\cal H}&=&\frac{{\bf E}^2}{\sqrt{1-\beta{\bf E}^2}\left(1+\sqrt{1-\beta{\bf E}^2}\right)}\nn\\
&=&\frac{{\bf D}^2}{\sqrt{1+\beta{\bf D}^2}+1}.
\eea
In view of \eq{2.2} and \eq{2.8}, we have the asymptotic expressions
\bea
{\cal H}({\bf x})&=&\frac{\left(\frac{q_i}{4\pi}\right)^2}{|\rr_i|^2\left(\sqrt{|\rr_i|^4+\beta \left(\frac{q_i}{4\pi}\right)^2}+|\rr_i|^2\right)},\quad |\rr_i|\ll1,\quad i=1,\dots,n,\\
{\cal H}({\bf x})&=& \frac1{2}\sum_{i,j=1}^n \frac{q_i q_j (\rr_i\cdot\rr_j)}{(4\pi)^2|\rr_i|^3|\rr_j|^3},\quad |{\bf x}|\gg1,
\eea
which extend the results of Born and Infeld \cite{B4}  in the radially symmetric situation with a singly-centered point electric charge.
Consequently, we see that the finiteness of the total energy follows as well,
\be
E=\int_{\bfR^3}{\cal H}\,\dd{\bf x}<\infty.
\ee
Furthermore, the expression \eq{2.4} gives us the result
\bea\lb{2.10}
\nabla\times{\bf E}
&=&-\frac1{4\pi}\sum_{i=1}^n \frac{q_i\rr_i}{|\rr_i|^3}\times \nabla\left(\frac1{\sqrt{1+\beta \left[\sum_{j=1}^n \frac{q_j\rr_j}{4\pi|\rr_j|^3}\right]^2}}\right)\nn\\
&=&-{\bf j}_m\equiv -{\bf j}^1_m-{\bf j}^2_m,
\eea
where
\be
{\bf j}^1_m=-\frac{\beta}{2(4\pi)^3\left(1+\beta \left[\sum_{l=1}^n \frac{q_l\rr_l}{4\pi|\rr_l|^3}\right]^2\right)^{\frac32}}\sum_{i,j,k=1}^n\frac{q_i q_j q_k\left(\rr_i\times [\rr_j+\rr_k]\right)}{|\rr_i|^3|\rr_j|^3|\rr_k|^3}
={\bf 0},
\ee
by applying the Jacobian identity
\be\lb{jacobian}
{\bf a}\times({\bf b}+{\bf c})+{\bf b}\times({\bf c}+{\bf a})+{\bf c}\times({\bf a}+{\bf b})={\bf 0},
\ee
for any three vectors in $\bfR^3$ in \eq{2.10}, and thus,
\be\lb{jm}
{\bf j}_m={\bf j}_m^2=\frac{3\beta}{2(4\pi)^3\left(1+\beta \left[\sum_{l=1}^n \frac{q_l\rr_l}{4\pi|\rr_l|^3}\right]^2\right)^{\frac32}}\sum_{i,j,k=1}^n\frac{q_i q_j q_k\left(\rr_j\cdot\rr_k\right)}{|\rr_i|^3|\rr_j|^3|\rr_k|^3}\,\rr_i\times\left(\frac{\rr_j}{|\rr_j|^2}+\frac{\rr_k}{|\rr_k|^2}\right).
\ee
From this expression, it is clear that ${\bf j}_m={\bf 0}$ in the single-point charge situation ($n=1$) and ${\bf j}_m\neq{\bf 0}$ in the multi-point charge situation  ($n\geq2$).
 In other words, in the Born--Infeld theory, the presence of multicentered electric point charges in equilibrium requires the presence of a static magnetic current to balance the
electric field generated.

As an illustration, we consider the case $n=2$. In this situation, we see that the magnetic current density \eq{jm} becomes
\be
{\bf j}_m=\frac{3\beta q_1 q_2}{(4\pi)^3(1+\beta{\bf D}^2)^{\frac32}|\rr_1|^3|\rr_2|^3}\left(\frac{(\rr_1\cdot\rr_2)}{|\rr_1|^2|\rr_2|^2}\left(\frac{q_1}{|\rr_1|}-\frac{q_2}{|\rr_2|}\right)+\frac{q_2}{|\rr_2|^3}-\frac{q_1}{|\rr_1|^3}\right)(\rr_1\times\rr_2).
\ee

Analogously but more easily, we can construct multicentered magnetostatic point charge sources because in this situation the magnetic field $\bf B$ given in \eq{1.8} with
\be\lb{2.13}
\rho_m({\bf x})=\sum_{i=1}^n g_i \delta(\rr_i),\quad  g_i\in\bfR\setminus\{0\},\quad i=1,\dots,n,
\ee
has the explicit form
\be\lb{2.14}
{\bf B}=\sum_{i=1}^n \frac{g_i\rr_i}{4\pi|\rr_i|^3}=-\nabla U_m,\quad U_m({\bf x})=\sum_{i=1}^n \frac{g_i}{4\pi|\rr_i|}.
\ee
Thus, by \eq{1.4}, the magnetic intensity field is given by
\be\lb{2.15}
{\bf H}=\frac{\bf B}{\sqrt{1+\beta{\bf B}^2}},
\ee
which is parallel to the $\bf E$ and $\bf D$ relation stated in \eq{2.4}. So $\bf H$ enjoys the same expressions as those for $\bf E$ with the replacement $q_i$ by $g_i$ ($i=1,\dots,n$).
As a result, the free magnetic charge density $\rho_m^{\mbox{\tiny free}}=\nabla\cdot{\bf H}$ leads to the total free magnetic charge 
\be
G_{\mbox{\tiny free}}=\int_{\bfR^3}\rho_m^{\mbox{\tiny free}}\,\dd{\bf x}=
\sum_{i=1}^n g_i \equiv G,\lb{2.16}
\ee
as that of the prescribed one. The magnetostatic Hamiltonian energy density is
\be\lb{2.17}
{\cal H}=\frac{{\bf B}^2}{\sqrt{1+\beta{\bf B}^2}+1},
\ee
which is analogous to \eq{2.8}.
Hence, in view of \eq{2.14} and \eq{2.17}, we see that the total energy of the magnetostatic multicentered point charge sources under consideration is finite as well. Moreover, from
\eq{2.14} and \eq{2.15}, we see that the electric current density given in \eq{1.6} assumes the form
\bea
{\bf j}_e&=&\nabla\times{\bf H}\nn\\
&=&-\frac{3\beta}{2(4\pi)^3\left(1+\beta \left[\sum_{l=1}^n \frac{g_l\rr_l}{4\pi|\rr_l|^3}\right]^2\right)^{\frac32}}\sum_{i,j,k=1}^n\frac{g_i g_j g_k\left(\rr_j\cdot\rr_k\right)}{|\rr_i|^3|\rr_j|^3|\rr_k|^3}\,\rr_i\times\left(\frac{\rr_j}{|\rr_j|^2}+\frac{\rr_k}{|\rr_k|^2}\right),
\eea
similar to the expression \eq{jm}, which is again nonvanishing except in the single-point charge situation. Thus we arrive at the same conclusion as before that, in the Born--Infeld theory,
in order to balance
a multicentered point magnetic charge distribution, an electric current density is needed.

\section{Dyonic point charges}\lb{sec4}
\setcounter{equation}{0}

We now consider dyonic point charge sources such that both electricity and magnetism are present and concurrently distributed following \eq{2.1} and \eq{2.13}.
First, using \eq{1.2}, multiplying both sides of \eq{1.3} by $\bf B$,  squaring it, and then squaring both sides of \eq{1.3}, we have
\bea
&& ({\bf B}\cdot{\bf D})^2(1-\beta[{\bf E}^2-{\bf B}^2+\kappa^2({\bf E}\cdot{\bf B})^2])=({\bf E}\cdot{\bf B})^2(1+\kappa^2{\bf B}^2)^2,\lb{3.1}\\
&&{\bf D}^2(1-\beta[{\bf E}^2-{\bf B}^2+\kappa^2({\bf E}\cdot{\bf B})^2])={\bf E}^2+\kappa^2 (2+\kappa^2{\bf B}^2)({\bf E}\cdot{\bf B})^2.\lb{3.2}
\eea
Solving for ${\bf E}^2$ and $({\bf E}\cdot{\bf B})^2$ in this system, we get
\bea
{\bf E}^2&=&\frac{(1+\beta{\bf B}^2)({\bf D}^2 +\kappa^2[2+\kappa^2{\bf B}^2][{\bf B}\times{\bf D}]^2)}{(1+\kappa^2{\bf B}^2)(1+\beta{\bf D}^2+\kappa^2{\bf B}^2+
\beta \kappa^2[{\bf B}\times{\bf D}]^2)},\lb{3.3}\\
({\bf E}\cdot{\bf B})^2&=&\frac{({\bf B}\cdot{\bf D})^2(1+\beta{\bf B}^2)}{(1+\kappa^2{\bf B}^2)(1+\beta{\bf D}^2+\kappa^2{\bf B}^2+
\beta \kappa^2[{\bf B}\times{\bf D}]^2)},\quad\lb{3.4}
\eea
where we have used the vector identity 
\be\lb{3.5}
({\bf B}\times{\bf D})^2={\bf B}^2{\bf D}^2-({\bf B}\cdot{\bf D})^2.
\ee
Applying \eq{3.3} and \eq{3.4} to \eq{1.3}, we arrive at
\bea
{\bf E}&=&\sqrt{1-2\beta s}\,{\bf D}-\kappa^2({\bf E}\cdot{\bf B}){\bf B}\nn\\
&=&\sqrt{1-2\beta s}\left({\bf D}-\frac{\kappa^2({\bf B}\cdot{\bf D})}{1+\kappa^2{\bf B}^2}\,{\bf B}\right)\nn\\
&=&\frac{\sqrt{1+\beta{\bf B}^2}\left([1+\kappa^2{\bf B}^2]{\bf D}-\kappa^2[{\bf B}\cdot{\bf D}]{\bf B}\right)}{\sqrt{1+\kappa^2{\bf B}^2}\sqrt{1+\beta{\bf D}^2+\kappa^2{\bf B}^2
+\beta\kappa^2 ({\bf B}\times {\bf D})^2}}.\lb{3.6}
\eea

For $n$ dyonic point charge sources located at ${\bf x}_1,\dots,{\bf x}_n$  with electric charges $q_1,\dots,q_n$ and magnetic charges $g_1,\dots,g_n$, the electric displacement field
$\bf D$ and magnetic field $\bf B$ are given by \eq{2.2} and \eq{2.14}, respectively. Hence the electric field $\bf E$ in \eq{3.6} has the properties 
\bea\lb{3.7}
{\bf E}({\bf x})&=&\frac{\sqrt{\beta}q_i }{\kappa\sqrt{\beta q_i^2+\kappa^2 g^2_i}}\frac{\rr_i}{|\rr_i|},\quad |\rr_i|\ll1,\quad i=1,\dots,n;\lb{3.7}\\
&=&\sum_{i=1}^n \frac{q_i\rr_i}{4\pi|\rr_i|^3},\quad
|{\bf x}|\gg1.\lb{3.8}
\eea
Therefore, as in the electrostatic situation, we see that the induced total free electric charge is the same as the prescribed one, $Q_{\mbox{\tiny free}}=\sum_{i=1}^n q_i=Q$.
Besides, from \eq{1.4} and \eq{3.6}, we obtain the magnetic intensity field
\bea\lb{3.9}
{\bf H}&=&\frac1{\cal D}{(1+\beta{\bf D}^2+\kappa^2(2{\bf B}^2 +2\beta[{\bf B}\times{\bf D}]^2+\beta[{\bf B}\cdot{\bf D}]^2)+\kappa^4([{\bf B}\cdot{\bf D}]^2+[1+\beta{\bf D}^2]	{\bf B}^4 )} {\bf B}\nn\\
&&-\frac{\kappa^2}{\cal D} (1+\beta {\bf B}^2)(1+\kappa^2{\bf B}^2)({\bf B}\cdot{\bf D}){\bf D},
\eea
expressed in terms of the prescribable fields $\bf B$ and $\bf D$ as before, where the quantity $\cal D$ in the denominator is given by
\be
{\cal D}={\sqrt{1+\beta {\bf B}^2}(1+\kappa^2{\bf B}^2)^{\frac32}\sqrt{1+\beta{\bf D}^2+\kappa^2{\bf B}^2+\beta\kappa^2({\bf B}\times{\bf D})^2}}.
\ee
The expression \eq{3.9} appears complicated. Nevertheless, we can insert \eq{2.2} and \eq{2.14} to obtain the asymptotic expressions
\bea\lb{3.11}
{\bf H}({\bf x})&=&\frac{\kappa g_i}{\sqrt{\beta}\sqrt{\beta q_i^2+\kappa^2 g^2_i}}\frac{\rr_i}{|\rr_i|},\quad |\rr_i|\ll1,\quad i=1,\dots,n;\lb{3.10}\\
&=&\sum_{i=1}^n \frac{g_i\rr_i}{4\pi|\rr_i|^3},\quad
|{\bf x}|\gg1,\lb{3.12}
\eea
which are similar to \eq{3.7} and \eq{3.8}. In particular, the total free magnetic charge is given by $G_{\mbox{\tiny free}}=\sum_{i=1}^n g_i=G$.
To compute the energy of such dyonic matter,  we note that the full Hamiltonian energy density is given by
\bea\lb{3.13}
{\cal H}&=&\frac{{\bf E}^2+\kappa^2({\bf E}\cdot{\bf B})^2}{\sqrt{1-2\beta s}}-\frac1{\beta}\left(1-\sqrt{1-2\beta s}\right)\nn\\
&=&\frac{{\bf E}^2+\kappa^2({\bf E}\cdot{\bf B})^2}{\sqrt{1-2\beta s}(1+\sqrt{1-2\beta s})}+\frac{{\bf B}^2}{1+\sqrt{1-2\beta s}},
\eea
where the quantity $s$ is defined by \eq{1.2}. Thus, inserting \eq{3.3} and \eq{3.4} into \eq{1.2}, we have
\be\lb{3.14}
s=\frac{{\bf D}^2-{\bf B}^2+\kappa^2([{\bf B}\times{\bf D}]^2-{\bf B}^4)}{2(1+\beta{\bf D}^2+\kappa^2{\bf B}^2+\beta\kappa^2[{\bf B}\times{\bf D}]^2)}.
\ee
As a consequence of \eq{3.3}, \eq{3.4}, and \eq{3.14}, we see that \eq{3.13} becomes
\bea
{\cal H}&=&\frac{{\bf B}^2 {\cal R}_1{\cal R}_2+(1+\beta{\bf B}^2)({\bf D}^2+\kappa^2[{\bf B}\times{\bf D}]^2)}{{\cal R}_1({\cal R}_1+{\cal R}_2)},\lb{3.15}\\
{\cal R}_1&=&\sqrt{(1+\beta{\bf B}^2)(1+\kappa^2{\bf B}^2)},\quad{\cal R}_2=\sqrt{1+\beta{\bf D}^2+\kappa^2{\bf B}^2+\beta\kappa^2({\bf B}\times{\bf D})^2}.\lb{3.16}
\eea
In view of \eq{2.2}, \eq{2.14}, \eq{3.15}, and \eq{3.16}, we get
\be
{\cal H}=\frac{\sqrt{\beta q_i^2+\kappa^2 g_i^2}}{4\pi\sqrt{\beta}\kappa|\rr_i|^2},\quad |\rr_i|\ll1; \quad {\cal H}=\frac{1}{2}\sum_{i,j=1}^n\frac{(\rr_i\cdot\rr_j)(q_i q_j+g_i g_j)}{(4\pi)^2|\rr_i|^3|\rr_j|^3},\quad
|{\bf x}|\gg1,
\ee
in leading orders. Thus the finiteness of the total energy of the static dyonic point charge source system follows.

Note that the discussion above relies on the assumption $\kappa>0$ and thus the limiting situation $\kappa=0$ needs to be treated separately which we now pursue. In fact, setting $\kappa=0$ in \eq{3.6}, we have
\be
{\bf E}=\frac{\sqrt{1+\beta{\bf B}^2} }{\sqrt{1+\beta{\bf D}^2}}\, {\bf D}.\lb{3.17}
\ee
Thus, by virtue of \eq{2.2} and \eq{2.14}, we have
\bea
{\bf E}({\bf x})&=&\frac{|g_i |\mbox{sgn}(q_i)}{4\pi}\frac{\rr_i}{|\rr_i|^3},\quad |\rr_i|\ll1,\quad i=1,\dots,n;\lb{3.18}\\
&=&\sum_{i=1}^n \frac{q_i\rr_i}{4\pi|\rr_i|^3},\quad
|{\bf x}|\gg1.\lb{3.19}
\eea
As a result, the total induced electric free charge is
\be\lb{3.20}
Q_{\mbox{\tiny free}}=\int_{\bfR^3}\nabla\cdot{\bf E}\,\dd{\bf x}=\lim_{r\to0,R\to \infty}\sum_{i=1}^n \int_{r<|\rr_i|<R}\nabla\cdot{\bf E}\,\dd {\bf x}=\sum_{i=1}^n q_i
-\sum_{i=1}^n|g_i|\mbox{sgn}(q_i).
\ee
In particular, in the interesting situation where $q_i$ and $g_i$ are of the same sign, $i=1,\dots,n$, then \eq{3.20} reads
\be
Q_{\mbox{\tiny free}}=\sum_{i=1}^n q_i
-\sum_{i=1}^n g_i=Q-G,
\ee
as defined in \eq{2.7}, although the free charge is seen to mix the prescribed ones. Analogous to \eq{3.17}, setting  $\kappa=0$ in \eq{3.9},  we have
\be
{\bf H}=\frac{\sqrt{1+\beta{\bf D}^2} }{\sqrt{1+\beta{\bf B}^2}}\, {\bf B}.\lb{3.22}
\ee
Hence, by the same computation, similar results hold for $\bf H$. For example, we see that the total free magnetic charge is given by the formula
\be\lb{3.23}
G_{\mbox{\tiny free}}=\int_{\bfR^3}\nabla\cdot{\bf H}\,\dd{\bf x}=\lim_{r\to0,R\to \infty}\sum_{i=1}^n \int_{r<|\rr_i|<R}\nabla\cdot{\bf H}\,\dd {\bf x}=\sum_{i=1}^n g_i
-\sum_{i=1}^n|q_i|\mbox{sgn}(g_i),
\ee
so that when all the electric and magnetic charges in pairs are of the same sign then $G_{\mbox{\tiny free}}=G-Q$. In this case, we have $Q_{\mbox{\tiny free}}=-G_{\mbox{\tiny free}}$,
in particular. The associated Hamiltonian energy density from \eq{3.15} with \eq{3.16} assumes the form
\be\lb{3.24}
{\cal H}=\frac{{\bf B}^2\sqrt{1+\beta{\bf D}^2}+{\bf D}^2\sqrt{1+\beta{\bf B}^2}}{\sqrt{1+\beta{\bf B}^2}+\sqrt{1+\beta{\bf D}^2}}.
\ee
Inserting \eq{2.2} and \eq{2.14} into \eq{3.24}, we obtain the asymptotic formulas
\bea
{\cal H}&=&\frac{\left(\frac {q_i}{4\pi}\right)^2\sqrt{r^4+\beta\left(\frac {g_i}{4\pi}\right)^2}+\left(\frac {g_i}{4\pi}\right)^2\sqrt{r^4+\beta\left(\frac {q_i}{4\pi}\right)^2}}{r^4\left(\sqrt{r^4+\beta\left(\frac {q_i}{4\pi}\right)^2}+\sqrt{r^4+\beta\left(\frac {g_i}{4\pi}\right)^2}\right)},\quad r=|\rr_i|\ll1,\quad i=1,\dots,n;\quad \lb{3.25}\\
&=&\frac{1}{2}\sum_{i,j=1}^n\frac{(\rr_i\cdot\rr_j)(q_i q_j+g_i g_j)}{(4\pi)^2|\rr_i|^3|\rr_j|^3},\quad |{\bf x}|\gg1.
\eea
From \eq{3.25}, we see that the energy diverges whenever there is a pair $q_i, g_i$ satisfying $q_i g_i\neq0$ for some $i=1,\dots,n$.

Although the dyonic matter at $\kappa=0$ is of infinite energy, the simplicity of its exact solution consisting of \eq{3.17} and \eq{3.22} enables us to gain insight into its
electric and magnetic current densities ${\bf j}_e$ and ${\bf j}_m$ in view of the non-conservativeness of ${\bf H}$ and ${\bf E}$ measured in terms of $\nabla\times {\bf H}$
and $\nabla\times{\bf E}$ given through \eq{1.5} and \eq{1.6}, respectively.
In fact,   inserting \eq{2.2} and \eq{2.14}, we get from \eq{3.17} the result
\bea\lb{jmm}
-{\bf j}_m=\nabla\times{\bf E}&=&{\sqrt{1+\beta{\bf B}^2} }\,\nabla\times \left(\frac{\bf D}{\sqrt{1+\beta{\bf D}^2}}\right)-\frac{\bf D}{\sqrt{1+\beta{\bf D}^2}}\times \nabla {\sqrt{1+\beta{\bf B}^2} }\nn\\
&=&-{\sqrt{1+\beta{\bf B}^2} }\,{\bf j}^2_m-{\bf j}_m^3,
\eea
where ${\bf j}_m^2$ is given in \eq{jm} and
\bea
&&{\bf j}_m^3=\frac{\bf D}{\sqrt{1+\beta{\bf D}^2}}\times \nabla {\sqrt{1+\beta{\bf B}^2} }\nn\\
&&=\frac\beta{2(4\pi)^3\sqrt{1+\beta{\bf D}^2}{\sqrt{1+\beta{\bf B}^2}}}\sum_{i,j,k=1}^n\frac{q_i g_j g_k}{|\rr_i|^3
|\rr_j|^3|\rr_k|^3}\rr_i\times \left(\rr_j+\rr_k-3(\rr_j\cdot\rr_k)\left(\frac{\rr_j}{|\rr_j|^2}+\frac{\rr_k}{|\rr_k|^2}\right)\right).\nn\\
\lb{3.28}
\eea
Consequently, ${\bf j}_m\neq{\bf 0}$ in the multicentered situation again. Similarly we also have $\nabla\times{\bf H}={\bf j}_e\neq{\bf 0}$ for the electric current density.

\section{Generalized formalism and the logarithmic model as an example}\lb{sec5}
\setcounter{equation}{0}

We may extend our construction to the nonlinear electrodynamics model governed by the generalized Lagrangian action density
\be\lb{4.1}
{\cal L}=f(s),
\ee
where $f(s)$ is a differentiable function satisfying $f(0)=0$ and $f'(s)=1$.  The equations \eq{1.5}--\eq{1.8} are still valid with \eq{1.3} and \eq{1.4} being replaced by
\bea
{\bf D}&=&f'(s)\left({\bf E}+{\kappa^2}({\bf E}\cdot{\bf B})\,{\bf B}\right), \lb{4.2} \\
{\bf H}&=&f'(s)\left({\bf B}-{\kappa^2}({\bf E}\cdot{\bf B})\,{\bf E}\right),\lb{4.3}
\eea
so that the nonlinear dielectrics and permeability coefficient matrix in \eq{1.9} is updated into the form
\be
 \Sigma({\bf E},{\bf B})\equiv\left(\begin{array}{cc}f'(s)(1+\kappa^4({\bf E}\cdot{\bf B})^2)& \kappa^2({\bf E}\cdot{\bf B})\\ \kappa^2({\bf E}\cdot{\bf B})&\frac1{f'(s)}\end{array}\right).
\ee

As a first concrete example of \eq{4.1}, consider the logarithmic model \cite{Soleng,Fe,AM,Gaete,K6}
\be\lb{4.5}
f(s)=-\frac1\beta\ln(1-\beta s).
\ee

In the electrostatic situation, we obtain from \eq{4.2} the simple relation
\be\lb{4.6}
{\bf D}\left(1-\frac\beta2 {\bf E}^2\right)={\bf E}.
\ee
Squaring \eq{4.6}, we have the solutions
\be\lb{4.7}
{\bf E}^2=\frac2\beta\left(1+\frac1{\beta{\bf D}^2}\pm\sqrt{\frac1{\beta {\bf D}^2}\left[2+\frac1{\beta{\bf D}^2}\right]}\right).
\ee
In order to ensure that the vector fields ${\bf D}$ and $\bf E$ are in the same direction, we need to choose the minus sign in \eq{4.7}, which ensures the bound
\be
{\bf E}^2<\frac2\beta. 
\ee
Thus, inserting this result, namely \eq{4.7} with the minus sign, into \eq{4.6},  we get
\be\lb{4.8}
{\bf E}=\frac{2{\bf D}}{1+\sqrt{1+2\beta{\bf D}^2}}.
\ee
As a consequence, the multicentered point charge source \eq{2.2} gives us the explicit expression 
\be\lb{4.9}
{\bf E}=
\sum_{i=1}^n \frac{q_i\rr_i}{2\pi|\rr_i|^3\left(1+\sqrt{1+2\beta \left(\sum_{j=1}^n \frac{q_j\rr_j}{4\pi|\rr_j|^3}\right)^2}\right)}.
\ee
Hence we have 
\be\lb{4.10}
{\bf E}({\bf x})=\frac{\sqrt{2}\,\mbox{sgn}(q_i)}{\sqrt{\beta}}\frac{\rr_i}{|\rr_i|},\quad |\rr_i|\ll1; \quad {\bf E}({\bf x})=\sum_{i=1}^n \frac{q_i\rr_i}{4\pi|\rr_i|^3},\quad
|{\bf x}|\gg1,
\ee
which is similar to \eq{2.5}. Therefore the total free electric charge is still given by \eq{2.7}. Moreover, the electric field \eq{4.9} is also non-conservative. Besides, with \eq{4.1}, we can
compute its associated Hamiltonian energy density to find
\be
{\cal H}=\frac{{\bf E}^2}{1-\frac\beta2{\bf E}^2}+\frac1\beta\ln\left(1-\frac\beta2{\bf E}^2\right).\lb{4.11}
\ee
From \eq{4.9}, \eq{4.10}, and \eq{4.11}, we see that the finiteness of the total energy of the multicentered point charge source considered follows.

In the magnetostatic situation, the relation \eq{4.3} and the multicentered point charge source \eq{2.14} give us
\be
{\bf H}=\frac{\bf B}{1+\frac\beta2{\bf B}^2}=\sum_{i=1}^n \frac{g_i\rr_i}{4\pi|\rr_i|^3\left(1+\frac\beta2\left[\sum_{j=1}^n\frac{g_j\rr_j}{4\pi |\rr_j|^3}\right]^2\right)},\lb{4.12}
\ee
which is direct and much simpler than but  different from \eq{4.9} locally, such that
\be\lb{4.13}
{\bf H}({\bf x})=\frac{8\pi}{{\beta}g_i}|\rr_i|\rr_i,\quad |\rr_i|\ll1; \quad {\bf H}({\bf x})=\sum_{i=1}^n \frac{g_i\rr_i}{4\pi|\rr_i|^3},\quad
|{\bf x}|\gg1.
\ee
Hence the total magnetic free charge is still determined by \eq{2.16} as before. Besides, the associated Hamiltonian energy density of the model \eq{4.5} reads
\be
{\cal H}=\frac1\beta\ln\left(1+\frac\beta2{\bf B}^2\right).\lb{4.14}
\ee
In view of \eq{2.14} and \eq{4.14}, it is clear that the total energy of the multicentered point magnetic charge source is also finite. In addition, the expression \eq{4.12} indicates that $\bf H$ is a non-conservative
field as expected.

We then consider a multicentered dyonic point charge source. For this purpose, with \eq{4.5}, we see that \eq{4.2} and \eq{4.3} become
\bea
(1-\beta s){\bf D}&=&{\bf E}+{\kappa^2}({\bf E}\cdot{\bf B})\,{\bf B}, \lb{4.15} \\
(1-\beta s){\bf H}&=&{\bf B}-{\kappa^2}({\bf E}\cdot{\bf B})\,{\bf E}.\lb{4.16}
\eea
 Multiplying \eq{4.15} by $\bf B$, squaring the result, and then squaring \eq{4.15} itself, and setting 
\be\lb{4.17}
a={\bf E}^2, \quad b=({\bf E}\cdot{\bf B})^2,
\ee
 to simplify notation, we obtain
\bea
({\bf B}\cdot{\bf D})^2\left(1-\frac\beta2(a-{\bf B}^2+\kappa^2 b)\right)^2&=&(1+\kappa^2{\bf B}^2)^2\,b,\lb{4.18}\\
{\bf D}^2\left(1-\frac\beta2(a-{\bf B}^2+\kappa^2 b)\right)^2&=&a+\kappa^2 (2+\kappa^2{\bf B}^2)b.\lb{4.19}
\eea
Combining \eq{4.18} and \eq{4.19}, we have
\be\lb{4.20}
b=\eta a,\quad \eta=\frac{({\bf B}\cdot{\bf D})^2}{{\bf D}^2+\kappa^2(2+\kappa^2{\bf B}^2)({\bf B}\times{\bf D})^2}.
\ee
Thus, inserting \eq{4.20} into \eq{4.19}, we arrive at the quadratic equation
\be
\frac{\beta^2}4(1+\kappa^2\eta)^2 a^2-\beta \left([1+\kappa^2\eta]\left[1+\frac\beta2 {\bf B}^2\right]+\frac{1+\kappa^2[2+\kappa^2 {\bf B}^2]\eta}{\beta{\bf D}^2}\right)a+
\left(1+\frac\beta2{\bf B}^2\right)^2=0,
\ee
in the quantity $a$.
This equation has two positive solutions. However, only the smaller solution gives rise to the correct direction match for the fields, which is
\bea\lb{4.22}
{\bf E}^2=a&=&\frac1{\beta(1+\kappa^2\eta)}\left(2+\beta{\bf B}^2+ \frac{2(1+\kappa^2[2+\kappa^2{\bf B}^2]\eta)}{\beta{\bf D}^2(1+\kappa^2\eta)}\right.\nn\\
&&\left.-2\sqrt{\frac{(1+\kappa^2[2+\kappa^2{\bf B}^2]\eta)}{\beta{\bf D}^2(1+\kappa^2\eta)}\left[2+\beta{\bf B}^2+ \frac{(1+\kappa^2[2+\kappa^2{\bf B}^2]\eta)}{\beta{\bf D}^2(1+\kappa^2\eta)}\right]}\right).
\eea
Therefore we have succeeded in expressing ${\bf E}^2$ in terms of ${\bf B}$ and ${\bf D}$ which are to be prescribed. Besides, by \eq{4.17} and \eq{4.20}, we have
\be
2s=a(1+\kappa^2\eta)-{\bf B}^2,
\ee
which depends on $\bf B$ and $\bf D$ only by \eq{4.22}. Multiplying \eq{4.15} by $\bf B$, we get
\be\lb{4.24}
{\bf E}\cdot{\bf B}=\frac{({\bf B}\cdot{\bf D})(1-\beta s)}{1+\kappa^2{\bf B}^2}.
\ee
Substituting \eq{4.24} into \eq{4.15} and using the result and \eq{4.24} in \eq{4.16}, we arrive at the solution
\bea
{\bf E}&=&(1-\beta s)\left({\bf D}-\frac{\kappa^2({\bf B}\cdot{\bf D})}{1+\kappa^2{\bf B}^2}{\bf B}\right),\lb{4.25}\\
{\bf H}&=&-\kappa^2(1-\beta s)\frac{({\bf B}\cdot{\bf D})}{1+\kappa^2{\bf B}^2}{\bf D}+\left(\frac1{1-\beta s}+\kappa^4(1-\beta s)\left[\frac{({\bf B}\cdot{\bf D})}{1+\kappa^2{\bf B}^2}\right]^2\right){\bf B},\lb{4.26}\\
s&=&\frac1\beta+\frac{1+\kappa^2(2+\kappa^2{\bf B}^2)\eta}{\beta^2{\bf D}^2(1+\kappa^2\eta)}\nn\\
&& -\frac1\beta \sqrt{\frac{(1+\kappa^2[2+\kappa^2{\bf B}^2]\eta)}{\beta{\bf D}^2(1+\kappa^2\eta)}\left[2+\beta{\bf B}^2+ \frac{(1+\kappa^2[2+\kappa^2{\bf B}^2]\eta)}{\beta{\bf D}^2(1+\kappa^2\eta)}\right]},\lb{4.27}\\
\eta&=&\frac{({\bf B}\cdot{\bf D})^2}{{\bf D}^2+\kappa^2(2+\kappa^2{\bf B}^2)({\bf B}\times{\bf D})^2},\lb{4.28}
\eea
in a summarized form expressed  in terms of the fields $\bf D$ and $\bf B$, given by \eq{2.2} and \eq{2.14}, explicitly. It is clear that the electrostatic solution \eq{4.8} is a special
case of \eq{4.25}--\eq{4.28} with setting ${\bf B}={\bf 0}$. Note that our method here relies on the assumption that ${\bf D}\neq{\bf 0}$.

Although \eq{4.25}--\eq{4.28} appear complicated, the special situation $\kappa=0$ is simple and deserves some consideration. In fact, in this situation, the relation \eq{4.2} leads to
\be
({\bf E}^2)^2-\frac2\beta\left(2+\beta{\bf B}^2+\frac2{\beta{\bf D}^2}\right){\bf E}^2+\frac1{\beta^2}(2+\beta{\bf B}^2)^2=0,
\ee
which gives rise to the relevant result
\be\lb{x4.31}
{\bf E}^2=\frac1\beta\left(2+\beta{\bf B}^2+\frac2{\beta {\bf D}^2}-2\sqrt{\frac1{\beta{\bf D}^2}\left[\frac1{\beta{\bf D}^2}+2+\beta{\bf B}^2\right]}\right).
\ee
Inserting this into \eq{4.15} (with $\kappa=0$), namely,
\be
\left(1-\frac\beta2[{\bf E}^2-{\bf B}^2]\right){\bf D}={\bf E},
\ee
we have 
\be\lb{x4.33}
{\bf E}=\frac{(2+\beta {\bf B}^2){\bf D}}{1+\sqrt{1+\beta {\bf D}^2(2+\beta {\bf B}^2)}},
\ee
which is relatively simpler. Hence, inserting \eq{2.2} and \eq{2.14}, we obtain the explicit expression for the electric field of the dyonic system:
\be\lb{x4.26}
{\bf E}=\left(\frac{2+\beta\left[\sum_{j=1}^n \frac{g_j\rr_j}{4\pi|\rr_j|^3}\right]^2}{1+\sqrt{1+\beta\left[\sum_{j=1}^n \frac{q_j\rr_j}{4\pi|\rr_j|^3}\right]^2 \left(2+\beta \left[\sum_{k=1}^n \frac{g_k\rr_k}{4\pi|\rr_k|^3}\right]^2  \right)}}\right) \sum_{i=1}^n \frac{q_i\rr_i}{4\pi|\rr_i|^3}.
\ee
Besides, with $\kappa=0$ in \eq{4.3} and comparing with \eq{x4.33}, we have the immediate result:
\be
{\bf H}=\frac{(1+\sqrt{1+\beta {\bf D}^2(2+\beta {\bf B}^2)}){\bf B}}{(2+\beta {\bf B}^2)}.
\ee
Thus,  inserting \eq{2.2} and \eq{2.14}, we obtain the magnetic intensity field
\be\lb{x4.27}
{\bf H}=\left(\frac{1+\sqrt{1+\beta\left[\sum_{j=1}^n \frac{q_j\rr_j}{4\pi|\rr_j|^3}\right]^2 \left(2+\beta \left[\sum_{k=1}^n \frac{g_k\rr_k}{4\pi|\rr_k|^3}\right]^2  \right)}}{2+\beta\left[\sum_{j=1}^n \frac{g_j\rr_j}{4\pi|\rr_j|^3}\right]^2}\right) \sum_{i=1}^n \frac{g_i\rr_i}{4\pi|\rr_i|^3},
\ee
explicitly as well.
 Consequently, we get the asymptotic expressions
\be\lb{4.31}
{\bf E}({\bf x})=\left|\frac{g_i}{q_i}\right|\frac{q_i\rr_i}{4\pi |\rr_i|^3},\quad |\rr_i|\ll1; \quad {\bf E}({\bf x})=\sum_{i=1}^n \frac{q_i\rr_i}{4\pi|\rr_i|^3},\quad
|{\bf x}|\gg1,
\ee
\be\lb{4.32}
{\bf H}({\bf x})=\left|\frac{q_i}{g_i}\right|\frac{g_i\rr_i}{4\pi |\rr_i|^3},\quad |\rr_i|\ll1; \quad {\bf H}({\bf x})=\sum_{i=1}^n \frac{g_i\rr_i}{4\pi|\rr_i|^3},\quad
|{\bf x}|\gg1.
\ee
Thus, as before, if $q_i$ and $g_i$ are of the same sign pairwise, then the total free electric and magnetic charges are simply given by $Q_{\mbox{\tiny free}}=Q-G$ and $G_{\mbox{\tiny free}}=G-Q$, respectively, which indicates that such
a dyonic matter distribution is of infinite energy, as we now examine. In fact, in this situation, the Hamiltonian energy density reads
\be\lb{4.33}
{\cal H}=\frac{{\bf E}^2}{1-\beta s}+\frac1\beta\ln(1-\beta s),
\ee
where the quantity $s$ given in \eq{4.27} with $\kappa=0$ gives us
\be\lb{4.34}
1-\beta s=\frac{2+\beta{\bf B}^2}{1+\sqrt{1+\beta{\bf D}^2(2+\beta{\bf B}^2)}}.
\ee
Using \eq{4.31} and \eq{4.34} in \eq{4.33}, we have
\be
{\cal H}=\frac{|q_i g_i|}{(4\pi)^2|\rr_i|^4},\quad |\rr_i|\ll1,\quad i=1,\dots,n,
\ee
within leading orders, which renders divergence of energy at the sites of the dyonic point charges.

We now consider the situation where $\kappa>0$.

Inserting \eq{2.2} and \eq{2.14} into \eq{4.25}--\eq{4.28}, we obtain 
\bea
{\bf E}({\bf x})&=&\left(\frac{\mbox{sgn}(q_i)}{\kappa |\rr_i|}-\frac{4\pi|\rr_i|}{\beta q_i}\right)\rr_i+\mbox{O}(|\rr_i|^4),\quad |\rr_i|\ll1,\\
{\bf H}({\bf x})&=&\frac{8\pi|\rr_i|}{\beta g_i}\,\rr_i+\mbox{O}(|\rr_i|^4),\quad |\rr_i|\ll1,
\eea
for $i=1,\dots,n$, and the behavior of $\bf E$ and $\bf H$ for $|{\bf x}|\gg1$ is still as described in \eq{4.31} and \eq{4.32}, respectively. As a consequence, we conclude that the
total free electric and magnetic charges of the dyonic point charge source coincide with the prescribed ones, $Q_{\mbox{\tiny free}}=Q$ and $G_{\mbox{\tiny free}}=G$. Hence the total
energy should be finite which we now confirm. Indeed, in this situation, the Hamiltonian energy density of the system becomes
\be\lb{4.38}
{\cal H}=\frac{1}{1-\beta s}({\bf E}^2+\kappa^2[{\bf E}\cdot{\bf B}]^2)+\frac1\beta\ln(1-\beta s),
\ee
where the quantity $s$ is given in \eq{4.27}. Inserting \eq{2.2} and \eq{2.14} into \eq{4.38}, we get the asymptotic expressions
\be
{\cal H}=\frac{|q_i|}{4\pi \kappa |\rr_i|^2},\quad |\rr_i|\ll1,\quad i=1,\dots,n;\quad {\cal H}=\frac{1}{2}\sum_{i,j=1}^n\frac{(\rr_i\cdot\rr_j)(q_i q_j+g_i g_j)}{(4\pi)^2|\rr_i|^3|\rr_j|^3},\quad |{\bf x}|\gg1.
\ee
Therefore the total energy is finite as expected.

It is clear that the fields $\bf E$  given in \eq{4.25} and \eq{x4.26} and $\bf H$ given in  \eq{4.26}, \eq{x4.27}, corresponding to the cases $\kappa>0$ and $\kappa=0$, respectively, are both non-conservative. Consequently, both magnetic and
electric currents are present.

\section{Other extended models}\lb{sec6}
\setcounter{equation}{0}
\setcounter{theorem}{0}
We first consider the exponential model \cite{H1,H2} defined by
\be\lb{5.1}
f(s)=\frac1\beta (\e^{\beta s}-1),\quad \beta>0,
\ee
which is the large-$p$ limit of the fractional-powered model \cite{Yang1,Yang2,Kr2}
\be\lb{5.2}
f_p(s)=\frac1\beta\left(\left[1+\frac{\beta s}p\right]^{p}-1\right), \quad p\geq1,\quad\beta>0.
\ee
Of course, $f_1(s)$ returns to the Maxwell theory. When $p\geq2$ is an integer,  \eq{5.2} is a polynomial model which allows a finite-energy electric point charge but rejects a magnetic one,
indicating the onset of an electromagnetic asymmetry \cite{Yang1}. The quadratic model $f_2(s)$ is of particular significance and interest in that it gives rise to a correct k-essence
cosmological evolution linking a radiation-dominated early-universe era to a dust-dominated final late-universe state \cite{Yang2}. Besides, it is shown in \cite{Yang1,Yang3} that the model
\eq{5.1} restores the electromagnetic symmetry that finite-energy electric,  magnetic, and dyonic point charges are allowed and that these finite-energy point charge sources may be used to construct charged black holes with relegated singularities. Thus, in this regard, it will be of interest to obtain multicentered point charge solutions for the model \eq{5.1} as well. A challenge
presented by \eq{5.1} is that the associated equations in some due steps cannot be solved explicitly so that an analytic approach has to be resorted to in order to acquire full information of the solutions. In this sense, the discussion here supplements what carried out in the previous sections, methodologically.

To proceed, we insert \eq{5.1} into \eq{4.2} and \eq{4.3} to get
\bea
{\bf D}&=&\e^{\beta s}\left({\bf E}+{\kappa^2}({\bf E}\cdot{\bf B})\,{\bf B}\right), \lb{5.3} \\
{\bf H}&=&\e^{\beta s}\left({\bf B}-{\kappa^2}({\bf E}\cdot{\bf B})\,{\bf E}\right).\lb{5.4}
\eea
Substituting \eq{1.2} into \eq{5.3}, we have
\bea
{\bf D}^2\e^{\beta {\bf B}^2}&=& \e^{\beta({\bf E}^2+\kappa^2[{\bf E}\cdot{\bf B}]^2)}({\bf E}^2+\kappa^2[2+\kappa^2{\bf B}^2][{\bf E}\cdot{\bf B}]^2),\lb{5.5}\\
\frac{({\bf B}\cdot{\bf D})^2\e^{\beta {\bf B}^2}}{1+\kappa^2{\bf B}^2}&=& \e^{\beta({\bf E}^2+\kappa^2[{\bf E}\cdot{\bf B}]^2)}(1+\kappa^2{\bf B}^2)({\bf E}\cdot{\bf B})^2.
\lb{5.6}
\eea
 Now multiplying \eq{5.5} by $\beta$ and subtracting by the $\beta\kappa^2$-multiple of \eq{5.6},
we have
\be
\beta\e^{\beta {\bf B}^2}\left({\bf D}^2-\frac{\kappa^2 ({\bf B}\cdot{\bf D})^2}{1+\kappa^2{\bf B}^2}\right)= \e^{\beta({\bf E}^2+\kappa^2[{\bf E}\cdot{\bf B}]^2)}\beta ({\bf E}^2+\kappa^2[{\bf E}\cdot{\bf B}]^2).
\ee
This equation is of the form
\be
\e^W W=\delta,
\ee
where the quantities $W$ and $\delta$ are given by
\bea
W&=&\beta({\bf E}^2+\kappa^2[{\bf E}\cdot{\bf B}]^2),\\
\delta&=&\beta \e^{\beta {\bf B}^2}\left({\bf D}^2-\frac{\kappa^2 ({\bf B}\cdot{\bf D})^2}{1+\kappa^2{\bf B}^2}\right)=\beta \e^{\beta {\bf B}^2}\frac{({\bf D}^2+
\kappa^2[{\bf B}\times{\bf D}]^2)}{1+\kappa^2{\bf B}^2},
\eea
which can be solved in terms of the Lambert $W$ function \cite{CG,Chow},
where $W(x)$ is defined by the relation $x=W\e^W$, so that $W(x)$ is analytic in the interval $x>-\frac1\e$ and enjoys the asymptotic expansions
\be\lb{5.11}
W(x)=\left\{\begin{array}{ll}\sum_{k=1}^\infty \frac{(-k)^{k-1}}{k!} x^k,&  x \mbox{ is around }0,\\
&\\
                                               \ln x-\ln\ln x+\frac{\ln\ln x}{\ln x}+\cdots,& x>3.\end{array}\right.
\ee
Hence
 we obtain
\be\lb{5.12}
\beta({\bf E}^2+\kappa^2[{\bf E}\cdot{\bf B}]^2)=W\left(\beta \e^{\beta {\bf B}^2}\frac{({\bf D}^2+
\kappa^2[{\bf B}\times{\bf D}]^2)}{1+\kappa^2{\bf B}^2}\right).
\ee
Thus,  we can express $s$ in terms of $\bf B$ and $\bf D$ only:
\be\lb{5.13}
s= -\frac{{\bf B}^2}2+\frac1{2\beta}W\left(\beta \e^{\beta {\bf B}^2}\frac{({\bf D}^2+
\kappa^2[{\bf B}\times{\bf D}]^2)}{1+\kappa^2{\bf B}^2}\right).
\ee
On the other hand,
multiplying \eq{5.3} by $\bf B$, we find
\be\lb{5.14}
{\bf E}\cdot{\bf B}=\frac{({\bf B}\cdot{\bf D})\e^{-\beta s}}{1+\kappa^2 {\bf B}^2}.
\ee
Thus we can resolve \eq{5.3} and \eq{5.4} to arrive at
\bea
{\bf E}&=&\e^{-\beta s}\,{\bf D}-\frac{\kappa^2({\bf B}\cdot{\bf D})}{1+\kappa^2 {\bf B}^2}\e^{-\beta s}\,{\bf B}, \lb{5.15} \\
{\bf H}&=&-\frac{\kappa^2({\bf B}\cdot{\bf D})}{1+\kappa^2 {\bf B}^2}\e^{-\beta s}\,{\bf D}+\left(\e^{\beta s}+\frac{\kappa^4({\bf B}\cdot{\bf D})^2}{(1+\kappa^2 {\bf B}^2)^2}
\e^{-\beta s}\right){\bf B},\lb{5.16}
\eea
where $s$ is given by \eq{5.13}. With $\bf D$ and $\bf B$ given by \eq{2.2} and \eq{2.14}, respectively,  the expressions \eq{5.15} and \eq{5.16} give us the exact multicentered dyonic
point charge solution for the exponential model \eq{5.1}. Since the associated Hamiltonian energy density now assumes the form
\bea
{\cal H}&=&\e^{\beta s}({\bf E}^2+\kappa^2[{\bf E}\cdot{\bf B}]^2)-\frac1\beta(\e^{\beta s}-1)\nn\\
&=&\frac1\beta\left(1-\e^{\beta s}[1-W]\right),
\eea
where $W$ and $s$ are defined by \eq{5.12} and \eq{5.13}, respectively, as functions of $\bf D$ and $\bf B$. Therefore, we can evaluate the total energy of a 
multicentered dyonic point charge system as before, which we omit. However, the conclusions we draw are the same: (i) For either the electric or magnetic point charge system, the
total energy is finite and the total free electric or magnetic charge is identical to the prescribed one. (ii) For a dyonic point charge system, the total energy is finite if and only if $\kappa>0$. In
the finite-energy situation, the total free electric and magnetic charges of the dyonic system coincide with the prescribed charges of the point charge source. (iii) In all these 
multicentered cases,
the electric field $\bf E$ and magnetic intensity field $\bf H$ are non-conservative.
 
We next consider the quadratic model \cite{De,K2007,GS,C2015,K2017} defined by
\be\lb{5.18}
f(s)=s+\alpha s^2,
\ee
where $\alpha>0$ is a parameter. Then \eq{4.2} reads
\be\lb{5.19}
{\bf D}=\left(1+\alpha \left[{\bf E}^2-{\bf B}^2+\kappa^2({\bf E}\cdot{\bf B})^2\right]\right)\left({\bf E}+{\kappa^2}({\bf E}\cdot{\bf B})\,{\bf B}\right).
\ee
From \eq{5.19}, we have
\bea
({\bf B}\cdot{\bf D})^2&=&\left(1+\alpha \left[{\bf E}^2-{\bf B}^2+\kappa^2({\bf E}\cdot{\bf B})^2\right]\right)^2\left(1+{\kappa^2}{\bf B}^2\right)^2 ({\bf E}\cdot{\bf B})^2,\lb{5.20}\\
{\bf D}^2&=&\left(1+\alpha \left[{\bf E}^2-{\bf B}^2+\kappa^2({\bf E}\cdot{\bf B})^2\right]\right)^2\left({\bf E}^2+\kappa^2[2+\kappa^2{\bf B}^2][{\bf E}\cdot{\bf B}]^2\right).\lb{5.21}
\eea
Eliminating the common factor in \eq{5.20} and \eq{5.21} and using the notation \eq{4.17}, we arrive at \eq{4.20} as before. Hence, in view of \eq{4.20}, we see that \eq{5.21} renders us
the cubic equation
\be\lb{5.22}
(1-\alpha {\bf B}^2+\alpha[1+\kappa^2\eta]a)^2(1+\kappa^2[2+\kappa^2{\bf B}^2]\eta)a={\bf D}^2.
\ee
For convenience, we normalize \eq{5.22} to get
\be
(\gamma+a)^2 a=\sigma^2;\quad \gamma=\frac{1-\alpha{\bf B}^2}{\alpha[1+\kappa^2\eta]},\quad \sigma^2=\frac{{\bf D}^2}{(\alpha[1+\kappa^2\eta])^2(1+\kappa^2[2+\kappa^2{\bf B}^2]\eta)},
\ee
so that
\be
{\bf E}^2=a=
\frac{((8\gamma^3+108\sigma^2+12\sqrt{12\gamma^3\sigma^2+81\sigma^4})^{\frac13}-2\gamma)^2}{
6(8\gamma^3+108\sigma^2+12\sqrt{12\gamma^3\sigma^2+81\sigma^4})^{\frac13}}.
\ee
As a consequence, we have
\be
{\bf E}^2+\kappa^2({\bf E}\cdot{\bf B})^2=(1+\kappa^2\eta)a,
\ee
such that multiplying \eq{5.19} by $\bf B$ leads us to the result
\be\lb{5.26}
{\bf E}\cdot{\bf B}=\frac{{\bf B}\cdot{\bf D}}{(1+\kappa^2{\bf B}^2)(1-\alpha{\bf B}^2+\alpha[1+\kappa^2\eta]a)}.
\ee
Substituting \eq{5.26} into \eq{5.19}, we obtain
\be\lb{5.27}
{\bf E}=\frac1{1-\alpha{\bf B}^2+\alpha(1+\kappa^2\eta)a}\left({\bf D}-\frac{\kappa^2({\bf B}\cdot{\bf D})}{1+\kappa^2{\bf B}^2}\,{\bf B}\right).
\ee
Besides, \eq{4.3} gives us
\bea\lb{5.28}
{\bf H}&=&\left(1+\alpha \left[{\bf E}^2-{\bf B}^2+\kappa^2({\bf E}\cdot{\bf B})^2\right]\right)\left({\bf B}-{\kappa^2}({\bf E}\cdot{\bf B})\,{\bf E}\right)\nn\\
&=&-\frac{\kappa^2({\bf B}\cdot{\bf D})}{(1+\kappa^2 {\bf B}^2)(1-\alpha{\bf B}^2+\alpha[1+\kappa^2\eta]a)}{\bf D}\nn\\
&&+\left(1-\alpha{\bf B}^2+\alpha[1+\kappa^2\eta]a+
\frac{\kappa^4({\bf B}\cdot{\bf D})^2}{(1+\kappa^2 {\bf B}^2)^2(1-\alpha{\bf B}^2+\alpha[1+\kappa^2\eta]a)}\right){\bf B}.
\eea
The expressions \eq{5.27} and \eq{5.28} give us the explicit solution to the multicentered dyonic point charge source problem for the quadratic model \eq{5.18} with the associated
Hamiltonian energy density
\bea\lb{5.29}
{\cal H}&=&(1+2\alpha s)({\bf E}^2+\kappa^2[{\bf E}\cdot{\bf B}]^2)-(s+\alpha s^2)\nn\\
&=&{\bf B}^2+(1+2\alpha {\bf B}^2)s+3\alpha s^2,\\
s&=&\frac12(1+\kappa^2\eta)a-\frac{{\bf B}^2}2.\lb{x5.29}
\eea
It can be examined that the same conclusions regarding the total energies, free electric and magnetic charges, and non-conservativeness of the electric and magnetic intensity fields of  multicentered electric, magnetic, and dyonic point charge
sources hold true in the current situation.

Finally, we sketch how to use the method here to obtain multicentered point charge systems in a generalized electrodynamics theory of the generic form \eq{4.1}. For this purpose, we first manipulate
\eq{4.2} to get
\bea
 ({\bf B}\cdot{\bf D})^2&=&(f'(s))^2(1+\kappa^2{\bf B}^2)^2 ({\bf E}\cdot{\bf B})^2,\lb{5.30}\\
{\bf D}^2&=&(f'(s))^2 ({\bf E}^2+\kappa^2[2+\kappa^2{\bf B}^2][{\bf E}\cdot{\bf B}]^2).\lb{5.31}
\eea
Eliminating the common factor $(f'(s))^2$ in \eq{5.30} and \eq{5.31} and using the notation \eq{4.17}, we get \eq{4.20}. Moreover, from \eq{5.30} and \eq{5.31}, we have
\bea\lb{5.32}
{\bf D}^2-\frac{\kappa^2({\bf B}\cdot{\bf D})^2}{1+\kappa^2{\bf B}^2}&=&(f'(s))^2({\bf E}^2+\kappa^2[{\bf E}\cdot{\bf B}]^2)\nn\\
&=&\left(f'\left(\frac12[1+\kappa^2\eta]a-\frac{{\bf B}^2}2\right)\right)^2(1+\kappa^2\eta)a.
\eea
Solving this equation, we get $a={\bf E}^2$ in terms of $\bf D$ and $\bf B$. Hence we find $b=({\bf E}\cdot{\bf B})^2$ also in terms of $\bf D$ and $\bf B$. Therefore
we obtain $s$ in the same sense. Now multiplying \eq{4.2} by $\bf B$, we have
\be\lb{5.33}
{\bf E}\cdot{\bf B}=\frac{{\bf B}\cdot{\bf D}}{f'(s)(1+\kappa^2{\bf B}^2)}.
\ee
Thus, inserting \eq{5.33} into \eq{4.2}, we arrive at
\be\lb{5.34}
{\bf E}=\frac1{f'(s)}\left({\bf D}-\frac{\kappa^2({\bf B}\cdot{\bf D})}{1+\kappa^2{\bf B}^2}\,{\bf B}\right).
\ee
Furthermore, using \eq{5.33} and \eq{5.34} in \eq{4.3}, we obtain
\be\lb{5.35}
{\bf H}=-\frac{\kappa^2({\bf B}\cdot{\bf D})}{f'(s)(1+\kappa^2{\bf B}^2)}\,{\bf D}+\left(f'(s)+\frac{\kappa^4({\bf B}\cdot{\bf D})^2}{f'(s)(1+\kappa^2{\bf B}^2)^2}\right){\bf B}.
\ee
The expressions \eq{5.34} and \eq{5.35}, along with the quantities $a$, $s$, and $\eta$ determined by \eq{5.32}, \eq{x5.29}, and \eq{4.20}, and the fields $\bf D$ and $\bf B$ 
prescribed by \eq{2.2} and \eq{2.14}, respectively, give the multicentered dyonic point charge system for the generalized nonlinear electrodynamics theory \eq{4.1}, with the
associated Hamiltonian energy density \cite{Yang3}:
\bea\lb{5.36}
{\cal H}&=&f'(s)({\bf E}^2+\kappa^2[{\bf E}\cdot{\bf B}]^2)-f(s)\nn\\
&=&\frac1{f'(s)}\left({\bf D}^2-\frac{\kappa^2({\bf B}\cdot{\bf D})^2}{1+\kappa^2{\bf B}^2}\right)-f(s)\nn\\
&=&\frac{{\bf D}^2+\kappa^2({\bf B}\times{\bf D})^2}{f'(s)(1+\kappa^2{\bf B}^2)}-f(s),
\eea
by subsequently applying \eq{5.33}, \eq{5.34}, and \eq{3.5} directly. In view of \eq{5.36}, we may readily evaluate the total energy of a multiply distributed dyonic point charge 
system in terms of the prescribed electric and magnetic data given for $\bf D$ and $\bf B$ and the coupling parameters $\beta$ and $\kappa$ as demonstrated in the examples shown earlier.

In the general setting, the computation may expectedly become rather cumbersome. Here we work out the simplest situation of a multicentered electric point charge source, which is of obvious basic interest. Thus, with ${\bf D}$ given by
\eq{2.2} and ${\bf B}={\bf0}$, we obtain from \eq{4.2} the results
\be\lb{5.38}
f'\left(\frac{{\bf E}^2}2\right){\bf E}={{\bf D}}.
\ee
This gives us the implicit equation
\be
\left(f'\left(\frac a2\right)\right)^2 a={\bf D}^2,
\ee
relating $a={\bf E}^2$ to ${\bf D}^2$. Resolving this equation, we have
\be\lb{5.40}
a=h({\bf D}^2).
\ee
Inserting \eq{5.40} into \eq{5.38} and using \eq{2.2}, we arrive at
\bea\lb{5.41}
{\bf E}&=&\frac{\bf D}{f'\left(\frac12 h({\bf D}^2)\right)}\nn\\
&=&\frac{\sum_{i=1}^n \frac{q_i\rr_i}{4\pi|\rr_i|^3}}{f'\left(\frac12 h\left(\left[\sum_{j=1}^n \frac{q_j\rr_j}{4\pi|\rr_j|^3}\right]^2\right)\right)}.
\eea
In this context, since $\nabla\times {\bf D}={\bf 0}$, we have
\bea
\nabla\times{\bf E}&=&-\sum_{i=1}^n \frac{q_i\rr_i}{4\pi|\rr_i|^3}\times \nabla \frac1{f'\left(\frac12 h\left(\left[\sum_{j=1}^n \frac{q_j\rr_j}{4\pi|\rr_j|^3}\right]^2\right)\right)}\nn\\
&=&\frac{3 f''\left(\frac12h({\bf D}^2)\right)h'({\bf D}^2)}{2(4\pi)^3\left(f'\left(\frac12h({\bf D}^2)\right)\right)^2}\sum_{i,j,k=1}^n\frac{q_i q_j q_k\left(\rr_j\cdot\rr_k\right)}{|\rr_i|^3|\rr_j|^3|\rr_k|^3}\,\rr_i\times\left(\frac{\rr_j}{|\rr_j|^2}+\frac{\rr_k}{|\rr_k|^2}\right),
\eea
where we have used the identity \eq{jacobian} again to do reduction. In particular, we see that $\bf E$ is non-conservative for {\em any} nonlinear electrodynamics, characterized by
the condition $f''(s)\not\equiv0$,  in the presence of a multicentered electric point charge source, as anticipated. Thus, the induced magnetic current ${\bf j}_m=-\nabla\times{\bf E}$ is
nontrivial. 

The analogous conclusion holds for the magnetic situation as well, more straightforwardly,  because \eq{4.3} gives us ${\bf H}=f'(s){\bf B}$ ($s=-\frac12 {\bf B}^2$), resulting in
\be
{\bf j}_e=\nabla \times{\bf H}=\frac12f''\left(-\frac12{\bf B}^2\right){\bf B}\times\nabla({\bf B}^2)\neq{\bf0},
\ee
in the multicentered point charge situation whenever $f''(s)\not\equiv0$ as before.

Summarizing our study in Sections \ref{sec5}--\ref{sec6}, we state

\begin{theorem}\lb{th2}
Consider the static solutions of the generalized Born--Infeld equations \eq{1.5}--\eq{1.8} defined by the Lagrangian action density \eq{4.1} in which the electric and magnetic fields $\bf E$ and $\bf B$ are related to the electric displacement
and magnetic intensity fields $\bf D$ and $\bf H$ through the constitutive equations \eq{4.2}--\eq{4.3}.

\begin{enumerate}
\item[(i)] In either the electric or magnetic situation subject to a multicentered  point charge source, the solution of the equations can be constructed explicitly and exactly and its induced electric or magnetic intensity field is non-conservative,
rendering the onset of a magnetic or electric current density, respectively, universally in all nonlinear models, characterized by $f''\neq0$.

\item[(ii)] In electric, magnetic, or dyonic situation, a precise construction of the solution of the equations describing a multicentered point charge source depends on
the specific structure of the model under consideration. For the logarithmic, exponential, and quadratic models, finite-energy and infinite-energy solutions can all be constructed
explicitly and exactly with a complete determination of the associated free charges and other local and global properties. These solutions all demonstrate the same
non-conservativeness properties of the induced electric and magnetic intensity fields, leading to the corresponding non-vanishing current densities, as in the classical case.

\end{enumerate}
\end{theorem}

Although in (i) of Theorem \ref{th2}, the statement regarding the non-conservativeness of the fields is made for the electric and magnetic charge situations, a similar conclusion may be
stated for the dyonic charge situation. To see this, we first resolve \eq{5.30} and \eq{5.31} to represent ${\bf E}^2$,  $({\bf E}\cdot{\bf B})^2$, and thus $s$ in terms of the quantities
${\bf D}^2$, ${\bf B}^2$, and $({\bf B}\cdot{\bf D})^2$. Then we obtain the expressions \eq{5.34} and \eq{5.35} for $\bf E$ and $\bf H$, respectively. Using
the property that $\nabla\times{\bf D}={\bf 0}$ and $\nabla\times{\bf B}={\bf 0}$ for ${\bf D}$ and ${\bf B}$ given in \eq{2.2} and \eq{2.14}, respectively, we see that the condition $f''(s)\not\equiv0$ again 
plays an essential role in making ${\bf E}$ and  ${\bf H}$ non-conservative.

\section{Continuous source situations}\lb{sec7}
\setcounter{equation}{0}
\setcounter{theorem}{0}

We now consider the continuous source situation. Assuming electrostatic limit with ${\bf B}={\bf0},{\bf H}={\bf0}$ and using the compressed notation $\rho=\rho_e$ and ${\bf j}={\bf j}_m$, the
 covariant
Maxwell equations \eq{1.5}--\eq{1.8} and the constitutive equations \eq{1.3}--\eq{1.4} become
\bea
&&\nabla\times{\bf E}=-{\bf j},\lb{7.1}\\
&&\nabla\cdot{\bf D}=\rho,\lb{7.2}\\
&&{\bf D}=\frac{\bf E}{\sqrt{1-\beta{\bf E}^2}},\lb{7.3}
\eea
respectively. Suggested by \eq{2.2}, we may represent the electric displacement field $\bf D$ as a conservative field,
\be\lb{7.4}
{\bf D}=\nabla u,
\ee
where $u$ is a scalar field over $\bfR^3$. In fact, by the Helmholtz decomposition \cite{Lamb}, we can represent $\bf D$
as ${\bf D}=\nabla u+\nabla\times {\bf A}$ for some scalar field $u$ and vector field $\bf A$. However, since $\bf A$ makes no presence in \eq{7.2},
we may set ${\bf A}={\bf 0}$, which returns to \eq{7.4} again.
 Then \eq{7.2} becomes a Poisson equation, $\Delta u=\rho$, such that we can formally express $u$ 
in the form of a Newton potential,
\be\lb{7.5}
u(\x)=(\Gamma * \rho)(\x)\equiv-\int_{\bfR^3} \frac{\rho({\bf y})}{4\pi|\x-{\bf y}|}\dd {\bf y},\quad \Gamma(\x)=-\frac1{4\pi|\x|},
\ee
as a convolution, when combining \eq{7.2} and \eq{7.4} (cf. \cite{Evans,GT}). It is a standard fact that $u(\x)$ enjoys the asymptotic estimate
\be\lb{7.6}
u(\x)=\mbox{O}(|\x|^{-\delta}),\quad |\x|\gg1,\quad \delta=\min\{1,\gamma-2\},
\ee
provided that $\rho(\x)=\mbox{O}(|\x|^{-\gamma})$ for $|\x|\gg1$ for some constant $\gamma>2$ and
\be\lb{7.7}
\int_{\bfR^3}|\rho(\x)|\,\dd \x<\infty.
\ee
See \cite{SY} for a discussion, for example. To ensure \eq{7.7}, it is sufficient to assume $\gamma>3$. With this assumption, \eq{7.6} reduces into
\be\lb{7.8}
u(\x)=\mbox{O}(|\x|^{-1}),\quad |\x|\gg1,
\ee
which agrees with that stated in \eq{2.2}. With \eq{7.4} and \eq{7.8}, we have ${\bf D}=\mbox{O}(|\x|^{-2})$ for $|\x|\gg1$, which agrees with
Coulomb's law.

With \eq{7.4} and \eq{7.5}, we resolve \eq{7.3} following the steps as in \eq{2.3} and \eq{2.4} to obtain
\be\lb{7.9}
{\bf E}=\frac{\nabla u}{\sqrt{1+\beta|\nabla u|^2}}=\frac{\nabla (\Gamma*\rho)}{\sqrt{1+\beta|\nabla(\Gamma*\rho)|^2}}.
\ee
This explicitly solves \eq{7.2}--\eq{7.3}, or, jointly, the equation
\be
\nabla\cdot\left(\frac{\bf E}{\sqrt{1-\beta{\bf E}^2}}\right)=\rho.
\ee

As before, it remains to examine \eq{7.1}. In fact, from \eq{7.9}, we have, with $\x=(x,y,z)$, the expression
\bea\lb{7.11}
\nabla\times{\bf E}&=&\frac{\beta (\nabla u\times\nabla|\nabla u|^2)}{2(1+\beta|\nabla u|^2)^{\frac32}}\nn\\
&=&\frac{\beta}{(1+\beta|\nabla u|^2)^{\frac32}}\nabla u\times(\nabla u\cdot\nabla u_x,\nabla u\cdot\nabla u_y,\nabla u\cdot\nabla u_z),
\eea
which is nontrivial in general. Interestingly, in the radially symmetric situation, $u=u(r)$, $r=|\x|$, we have
\bea
\nabla u&=&\frac{u'(r)}r{\bf x},\\
(\nabla u\cdot\nabla u_x,\nabla u\cdot\nabla u_y,\nabla u\cdot\nabla u_z)&=&\frac{u'(r)u''(r)}r{\bf x}.
\eea
Hence in this situation the right-hand side of \eq{7.11} vanishes. In other words, in the radially symmetric situation, the magnetic current density
${\bf j}={\bf j}_m$ in \eq{7.1} is absent, although in a general situation, it is present.

The magnetostatic situation is similar but simpler, and thus omitted. The conclusion is that an electric current density ${\bf j}_e$ is 
generally needed to main a nonradially 
symmetric continuous magnetic charge distribution $\rho_m$, and, ${\bf j}_e$ vanishes when $\rho_m$ is radially symmetric.

In the dyonic source situation, the equations \eq{1.7} and \eq{1.8} lead to
\be\lb{7.14}
{\bf D}=\nabla (\Gamma*\rho_e),\quad {\bf B}=\nabla(\Gamma*\rho_m),
\ee
such that the unknown fields are $\bf E$ and $\bf H$. In fact, the equations \eq{1.3}, \eq{1.4}, \eq{1.7}, and \eq{1.8} jointly and
straightforwardly give us the  equations
\bea
&&\nabla\cdot\left(\frac{{\bf E}+\kappa^2({\bf E}\cdot\nabla(\Gamma*\rho_m))\nabla(\Gamma*\rho_m)}{\sqrt{1-\beta \sigma}}\right)=\rho_e,\lb{7.15}\\
&&\nabla\cdot\left(\sqrt{1-\beta\sigma}\,{\bf H}+\kappa^2({\bf E}\cdot \nabla(\Gamma*\rho_m)){\bf E}\right)=\rho_m,\lb{7.16}
\eea
governing $\bf E$ and $\bf H$, where
\be\lb{7.17}
\sigma=2s={\bf E}^2-|\nabla(\Gamma*\rho_m)|^2+\kappa^2 ({\bf E}\cdot \nabla(\Gamma*\rho_m))^2.
\ee
Inserting \eq{7.14} into \eq{3.6} and \eq{3.9}, we obtain the solutions to \eq{7.15}--\eq{7.17}. Hence we have solved the full Born--Infeld
dyonic equations comprised of \eq{1.3}--\eq{1.8}, exactly and explicitly.

Although \eq{7.15}--\eq{7.17} look complicated, the situation $\kappa=0$ greatly simplifies the system and is worth noting. In this situation, we
may manipulate \eq{1.3}
and \eq{1.4} to get
\be\lb{7.18}
{\bf D}={\bf E}\frac{\sqrt{1-\beta{\bf H}^2}}{\sqrt{1-\beta{\bf E}^2}},\quad {\bf B}={\bf H}\frac{\sqrt{1-\beta{\bf E}^2}}{\sqrt{1-\beta{\bf H}^2}}.
\ee
Thus, applying \eq{1.7} and \eq{1.8}, we obtain the following general dyonic matter source equations
\be\lb{7.19}
\nabla\cdot\left({\bf E}\frac{\sqrt{1-\beta{\bf H}^2}}{\sqrt{1-\beta{\bf E}^2}}\right)=\rho_e,\quad 
\nabla\cdot\left({\bf H}\frac{\sqrt{1-\beta{\bf E}^2}}{\sqrt{1-\beta{\bf H}^2}}\right)=\rho_m.
\ee
In particular, when $\bf E$ and $\bf H$ are both conservative such that $\nabla\times{\bf E}={\bf 0}$ and $\nabla\times{\bf H}={\bf 0}$,
then they are generated by some real scalar potential fields $\phi$ and $\psi$ with ${\bf E}=\nabla \phi$ and ${\bf H}=\nabla \psi$. In this situation,
the system \eq{7.19} becomes a familiar one \cite{Yang3},
 \be\lb{7.20}
\nabla\cdot\left(\nabla\phi\frac{\sqrt{1-\beta|\nabla\psi|^2}}{\sqrt{1-\beta|\nabla\phi|^2}}\right)=\rho_e,\quad 
\nabla\cdot\left(\nabla\psi\frac{\sqrt{1-\beta|\nabla\phi|^2}}{\sqrt{1-\beta|\nabla\psi|^2}}\right)=\rho_m.
\ee
However, we now know that, in a nonradially symmetric situation such that $\bf E$ and $\bf H$ fail to be conservative, the scalar-field reduction of the
problem from \eq{7.19} into \eq{7.20} may {\em not} go through as described.

We now consider the situation $\kappa>0$ of the system of equations \eq{1.3}--\eq{1.8}. First, from \eq{1.4}, we have
\bea
{\bf H}^2(1-\beta[{\bf E}^2-{\bf B}^2+\kappa^2({\bf E}\cdot{\bf B})^2])&=&{\bf B}^2-\kappa^2({\bf E}\cdot{\bf B})^2(2-\kappa^2{\bf E}^2),\\
({\bf E}\cdot{\bf H})^2(1-\beta[{\bf E}^2-{\bf B}^2+\kappa^2({\bf E}\cdot{\bf B})^2])&=&({\bf E}\cdot{\bf B})^2(1-\kappa^2{\bf E}^2)^2.
\eea
Resolving these equations for ${\bf B}^2$ and $({\bf E}\cdot{\bf B})^2$, we get
\bea
{\bf B}^2&=&\frac{(1-\beta{\bf E}^2)({\bf H}^2-\kappa^2 [{\bf E}\times{\bf H}]^2[2-\kappa^2{\bf E}^2])}{(1-\kappa^2{\bf E}^2)
(1-\beta{\bf H}^2-\kappa^2{\bf E}^2+\beta\kappa^2[{\bf E}\times{\bf H}]^2)},\\
({\bf E}\cdot{\bf B})^2&=&\frac{(1-\beta{\bf E}^2)({\bf E}\cdot{\bf H})^2}{(1-\kappa^2{\bf E}^2)(1-\beta{\bf H}^2-\kappa^2{\bf E}^2+\beta\kappa^2[{\bf E}\times{\bf H}]^2)},\lb{7.24}
\eea
which results in the expression
\be\lb{7.25}
s=\frac{{\bf E}^2-{\bf H}^2+\kappa^2([{\bf E}\times{\bf H}]^2-{\bf E}^2)}{2(1-\beta{\bf H}^2-\kappa^2{\bf E}^2+\beta\kappa^2[{\bf E}\times{\bf H}]^2)}.
\ee
On the other hand, we may rewrite \eq{1.3} and \eq{1.4} as
\bea
{\bf D}&=&\frac{(1+\kappa^4 [{\bf E}\cdot{\bf B}]^2){\bf E}}{\sqrt{1-2\beta s}}+\kappa^2({\bf E}\cdot{\bf B}){\bf H},\lb{7.26}\\
{\bf B}&=&\sqrt{1-2\beta s}\,{\bf H}+\kappa^2({\bf E}\cdot{\bf B}){\bf E}.\lb{7.27}
\eea
Inserting \eq{7.24} and \eq{7.25} into \eq{7.26} and \eq{7.27}, we see that \eq{1.7} and \eq{1.8} become
\bea
&&\nabla\cdot\left(\frac1{1-\kappa^2{\bf E}^2}\left(\frac1{\cal R}+\kappa^4({\bf E}\cdot{\bf H})^2{\cal R}\right){\bf E}+\kappa^2({\bf E}\cdot{\bf H}){\cal R}\,{\bf H}\right)=\rho_e,\lb{7.28}\\
&&\nabla\cdot\left(\kappa^2({\bf E}\cdot{\bf H}){\cal R}\,{\bf E}+(1-\kappa^2{\bf E}^2){\cal R}\,{\bf H}\right)=\rho_m,\lb{7.29}
\eea
where 
\be\lb{7.30}
{\cal R}=\frac{\sqrt{1-\beta{\bf E}^2}}{\sqrt{(1-\kappa^2{\bf E}^2)(1-\beta{\bf H}^2+\kappa^2(\beta[{\bf E}\times{\bf H}]^2-{\bf E}^2))}}.
\ee
In the limiting situation when $\kappa=0$, these equations return to \eq{7.19}, of course. 

An advantage of the system of equations, \eq{7.28} and \eq{7.29}, with \eq{7.30}, over \eq{7.15} and \eq{7.16}, with \eq{7.17}, is that the electric and magnetic charge densities, $\rho_e$ and $\rho_m$, both appear as nonhomogeneous terms on
the right-hand sides of the equations, but make no presence on the left-hand sides, where only the unknowns $\bf E$ and $\bf H$ are present.

For the generalized Born--Infeld model \eq{4.1}, we see that the electrostatic situation and the constitutive equation \eq{5.38} give us the equation
\be
\nabla\cdot\left(f'\left(\frac{{\bf E}^2}2\right){\bf E}\right)=\rho_e,
\ee
whose solution following from $\bf D$ given in \eq{7.14} and the first line in \eq{5.41} reads
\be
{\bf E}=\frac{\nabla(\Gamma *\rho_e)}{f'\left(\frac12 h(|\nabla(\Gamma *\rho_e)|^2)\right)}.
\ee
Hence, with $u=(\Gamma*\rho_e)$, we have
\be
\nabla\times{\bf E}=\frac{f''\left(\frac12 h(|\nabla u|^2)\right)h'(|\nabla u|^2)}{\left(f'\left(\frac12 h(|\nabla u|^2)\right)\right)^2}\,\nabla u\times(\nabla u\cdot\nabla u_x,\nabla u\cdot\nabla u_y,\nabla u\cdot\nabla u_z),
\ee
as that in \eq{7.11}. So, again, $\nabla\times{\bf E}$ vanishes when either $u$ is radially symmetric, or $f''(s)\equiv0$, as in linear electrodynamics.
 
In the fully dyonic situation, from \eq{4.3}, we have
\bea
{\bf H}^2&=&(f'(s))^2({\bf B}^2+\kappa^2({\bf E}\cdot{\bf B})^2(\kappa^2{\bf E}^2-2)),\lb{7.34}\\
({\bf E}\cdot{\bf H})^2&=&(f'(s))^2({\bf E}\cdot{\bf B})^2(1-\kappa^2{\bf E}^2)^2.\lb{7.35}
\eea
With $s=\frac12({\bf E}^2-{\bf B}^2+\kappa^2[{\bf E}\cdot{\bf B}]^2)$ in \eq{7.34} and \eq{7.35}, we can solve for ${\bf B}^2$ and $({\bf E}\cdot{\bf B})^2$ to get these quantities in terms of ${\bf E}^2$, ${\bf H}^2$, and $({\bf E}\cdot{\bf H})^2$. Consequently, we obtain from \eq{7.35} the expression
\be
({\bf E}\cdot{\bf B})=\frac{({\bf E}\cdot{\bf H})}{f'(s)(1-\kappa^2{\bf E}^2)}.
\ee
 Inserting this result back into \eq{4.2} and \eq{4.3}, we arrive at
\bea
{\bf D}&=&\left(f'(s)+\frac{\kappa^4({\bf E}\cdot{\bf H})^2}{f'(s)(1-\kappa^2{\bf E}^2)^2}\right){\bf E}+\frac{\kappa^2({\bf E}\cdot{\bf H})}{f'(s)(1-\kappa^2{\bf E}^2)}\,{\bf H},\lb{7.37}\\
{\bf B}&=&\frac{\kappa^2({\bf E}\cdot{\bf H})}{f'(s)(1-\kappa^2{\bf E}^2)}\,{\bf E}+\frac{\bf H}{f'(s)}.\lb{7.38}
\eea
Thus, with \eq{7.37} and \eq{7.38}, we see that the equations \eq{1.7} and \eq{1.8} become
\bea
&&\nabla\cdot\left(\left(f'(s)+\frac{\kappa^4({\bf E}\cdot{\bf H})^2}{f'(s)(1-\kappa^2{\bf E}^2)^2}\right){\bf E}+\frac{\kappa^2({\bf E}\cdot{\bf H})}{f'(s)(1-\kappa^2{\bf E}^2)}\,{\bf H}\right)=\rho_e,\lb{7.39}\\
&&\nabla\cdot\left(\frac{\kappa^2({\bf E}\cdot{\bf H})}{f'(s)(1-\kappa^2{\bf E}^2)}\,{\bf E}+\frac{\bf H}{f'(s)}\right)=\rho_m.\lb{7.40}
\eea
This system is the generalized form of the system of equations \eq{7.28} and \eq{7.29}.

To solve these coupled equations, rewrite \eq{7.14} as ${\bf D}=\nabla u$ and ${\bf B}=\nabla v$ and insert these into \eq{5.34} and \eq{5.35} to get
the exact and explicit solution
\bea
{\bf E}&=&\frac1{f'(s)}\left(\nabla u-\frac{\kappa^2(\nabla u\cdot\nabla v)}{1+\kappa^2|\nabla v|^2}\,\nabla v\right),\lb{7.41}\\
{\bf H}&=&-\frac{\kappa^2(\nabla u\cdot\nabla v)}{f'(s)(1+\kappa^2|\nabla v|^2)}\,\nabla u+\left(f'(s)+\frac{\kappa^4(\nabla u\cdot\nabla v)^2}{f'(s)(1+\kappa^2|\nabla v|^2)^2}\right)\nabla v,\lb{7.42}
\eea
to the system of equations \eq{7.39} and \eq{7.40}.

Note that, from \eq{5.30} and \eq{5.31}, we can solve for ${\bf E}^2$ and $({\bf E}\cdot{\bf B})^2$ in terms of ${\bf D}^2$, ${\bf B}^2$, and
$({\bf B}\cdot{\bf D})^2$. Hence the quantity $s$ may be regarded as expressed in terms of ${\bf D}^2=|\nabla u|^2$, ${\bf B}^2=|\nabla v|^2$, and
$({\bf B}\cdot{\bf D})^2=(\nabla u\cdot\nabla v)^2$ already, say,
\be\lb{7.43}
s=s\left({\bf D}^2,{\bf B}^2,({\bf B}\cdot {\bf D})^2\right);\quad s\to0\mbox{ as }{\bf D},{\bf B}\to {\bf 0},
\ee
which renders \eq{7.41} and \eq{7.42} explicit. Hence, as before, in the radially symmetric situation where $u$ and $v$ are functions of $r=|{\bf x}|$ only, we have
$\nabla\times {\bf E}={\bf 0}$ and $\nabla\times{\bf H}={\bf0}$, in view of \eq{7.41} and \eq{7.42}. Otherwise, this 
conservativeness condition on $\bf E$ and $\bf H$ is violated in general.

We may summarize the study of this section as follows.

\begin{theorem}
The methods presented in this work for an explicit construction of the exact solutions to the classical and generalized Born--Infeld equations
with multicentered point charge sources may be carried over to obtain solutions with arbitrarily distributed continuous charge sources.
\begin{enumerate}
\item[(i)] In the classical model situation, electrostatic, magnetostatic, and dyonic static solutions are obtained with the properties that
the associated electric, magnetic, and electric and magnetic fields are conservative in the radially symmetric situation but not so in general
in a nonradially symmetric situation, leading to the appearance of magnetic, electric, and electric and magnetic currents, respectively.

\item[(ii)] In the generalized model situation, the same conclusions as those in the classical model are also valid. 

\item[(iii)] In all situations, the induced electric field $\bf E$ and magnetic intensity field $\bf H$ enjoy the same asymptotic estimates as those
of the electric displacement field $\bf D$ and magnetic field $\bf B$ which are determined by
the prescribed electric charge density $\rho_e$ and
magnetic charge density $\rho_m$ through \eq{1.7} and \eq{1.8}, or \eq{7.14},
provided that $\rho_e$ and $\rho_m$  satisfy the condition
\be \lb{7.45}
\rho_e({\bf x})=\mbox{\rm O}(|{\bf x}|^{-\gamma}),\quad \rho_m({\bf x})=\mbox{\rm O}(|{\bf x}|^{-\gamma}),\quad |{\bf x}|\gg1,\quad\gamma>3.
\ee
More precisely, under the condition \eq{7.45}, we have
\be\lb{7.44}
{\bf E},{\bf D}=\mbox{\rm O}(|{\bf x}|^{-2}),\quad {\bf H}, {\bf B}=\mbox{\rm O}(|{\bf x}|^{-2}),\quad |{\bf x}|\gg1,
\ee
which are consistent with Coulomb's law. 
Consequently, all such solutions are of finite total energies and total charges.
\end{enumerate}
\end{theorem}

Here we only need to establish (iii) since (i) and (ii) are already shown in detail earlier. In fact, from \eq{7.14} and \eq{7.45}, we know that
$u=\Gamma*\rho_e$ and $v=\Gamma*\rho_m$ satisfy the estimates \cite{SY}
\be\lb{7.46}
u({\bf x}),v({\bf x})=\mbox{O}(|{\bf x}|^{-1}),\quad |{\bf x}|\gg1.
\ee
Thus \eq{7.44} follows from using \eq{7.46} in ${\bf D}=\nabla u, {\bf B}=\nabla v$, and \eq{7.41}--\eq{7.43}.
\medskip

Note that, since the solutions are all explicitly obtained, their regularity or smoothness is well exhibited by that of the source terms, 
$\rho_e$ and $\rho_m$, through $\Delta u=\rho_e$ and $\Delta v=\rho_m$, in view of the standard linear elliptic regularity theory \cite{Evans,GT}.

\section{Conclusions}

In this work, we have explicitly constructed exact solutions to the static Born--Infeld equations of nonlinear electrodynamics and its various extensions realizing
multicentered electric, magnetic, and dyonic point charge source distributions and have described their characteristic properties including their free electric and magnetic charges arising
as global quantities given by the induced electric and magnetic intensity fields determined by prescribed point charges given locally. The electric and magnetic solutions all carry finite energies
but the dyonic solutions carry either a finite or infinite energy, depending on whether an underlying electromagnetic coupling constant  in the theory is non-vanishing or vanishing.
Moreover, a universal feature present in all these multicentered solutions is that the induced electric and magnetic intensity fields for the electric, magnetic, and dyonic point charge situations
are non-conservative in general, leading to the onsets of magnetic and electric currents, respectively, as a consequence of the nonlinearity imposed that couples electricity and
magnetism together by the nonlinear constitutive equations formulated in the theory. Furthermore, we have shown that the multicentered constructions and
conclusions can be carried over to the situation of continuously distributed charge source problems.
\medskip

The author thanks an anonymous referee whose questions and suggestions led to an improvement of the presentation of the results of this work.
The author also thanks Shouxin Chen for stimulating discussions regarding Section \ref{sec7}.
\medskip

{\bf Data availability statement}: The data that supports the findings of this study are
available within the article.


\begin{thebibliography}{99}

\bibitem{Sch1}
J. Schwinger, Magnetic charge and quantum field theory, {\em Phys. Rev.} {\bf144} (1966) 1087--1093.

\bibitem{Sch2}
J. Schwinger, A magnetic model of matter, {\em Science} {\bf165} (1969) 757--761.

\bibitem{Sch3}
J. Schwinger, Magnetic charge and the charge quantization condition, {\em Phys. Rev.} D {\bf12} (1975) 3105--3111.

\bibitem{Z}
D. Zwanziger, Quantum field theory of particles with both electric and magnetic charges, {\em Phys. Rev.} {\bf176} (1968) 1489--1495.

\bibitem{Jackson}
J . D. Jackson, {\em Classical Electrodynamics}, 2nd ed., Wiley, New York, 1975.

\bibitem{Bla}
M. Blagojevit and P. Senjanovic, The quantum field theory of electric and magnetic charge, {\em Phys. Reports} {\bf157} (1988) 233--346.

\bibitem{Sin}
D. Singleton, Magnetic charge as a ``hidden" gauge symmetry, {\em Internat. J. Theoret. Phys.} {\bf34} (1995) 37--46.

\bibitem{Singleton}
D. Singleton, Electromagnetism with magnetic charge and two photons, {\em Amer. J. Phys.} {\bf64} (1996) 452--458.



\bibitem{Mig}
J. A. Mignaco, Electromagnetic duality, charges, monopoles, topology, {\em Brazilian J. Phys.} {\bf31} (2001) 235--246.

\bibitem{Milton}
K. A. Milton, Theoretical and experimental status of magnetic monopoles, {\em Repts. Prog. Phys.} {\bf69} (2006) 1637--1711.

\bibitem{Dirac}
P. Dirac, Quantised singularities in the electromagnetic field, {\em Proc. Roy. Soc.}  A {\bf133}
(1931) 60--72.

\bibitem{B1}
M. Born, Modified field equations with a finite radius of the electron,
{\em Nature} {\bf 132} (1933) 282.

\bibitem{B2}
M. Born, On the quantum theory of the electromagnetic field, {\em Proc. Roy. Soc.} A
{\bf 143} (1934) 410--437.


\bibitem{B3}
M. Born and L. Infeld, Foundation of the new field theory, {\em Nature} {\bf132} (1933) 1004.

\bibitem{B4}
M. Born and L. Infeld, Foundation of the new field theory, {\em Proc. Roy. Soc.}
A {\bf 144} (1934) 425--451.

\bibitem{FT}
E. S. Fradkin and A. A. Tseytlin, Non-linear electrodynamics from quantized strings, {\em Phys. Lett.}  B. {\bf163} (1985)123--130.

\bibitem{Ts2}
 A. A. Tseytlin, On non-abelian generalisation of Born--Infeld action
in string theory, {\em Nucl. Phys.} B {\bf501} (1997) 41--52.

\bibitem{Tsey}
A. A. Tseytlin, Born--Infeld action, supersymmetry and string theory,  {\em The Many Faces of the Superworld},  pp. 417--452, World Scientific, Singapore, 2000.


\bibitem{CM}
C. G. Callan Jr. and J. M. Maldacena, 
Brane dynamics from the Born--Infeld action, {\em Nucl. Phys.} B {\bf513} (1998) 198--212.

\bibitem{Gibbons}
G. W. Gibbons,  Born--Infeld particles and Dirichlet $p$-branes, {\em Nucl. Phys.} B {\bf514} (1998) 603--639.

\bibitem{Ts1}
A. A. Tseytlin, Self-duality of Born--Infeld action and Dirichlet 3-brane of type IIB superstring theory, {\em Nucl. Phys.} B {\bf469} 
(1996) 51--67.

\bibitem{AG1}
E. Ay\'{o}n--Beato and A. Garc\'{i}a, The Bardeen model as a nonlinear magnetic monopole, {\em Phys. Lett.} B {\bf493} (2000)
149--152.

\bibitem{AG2}
E. Ay\'{o}n--Beato and A. Garc\'{i}a, Regular black hole in general relativity coupled to nonlinear electrodynamics, 
{\em Phys. Rev. Lett.} {\bf 80} (1998) 5056--5059.

\bibitem{K1}
S. I. Kruglov, Born--Infeld-type electrodynamics and magnetic black holes,
{\em Ann. Phys.} {\bf383} (2017) 550--559.


\bibitem{K2}
S. I. Kruglov,
Dyonic black holes in framework of Born--Infeld-type electrodynamics,
{\em Gen. Relat. Grav.} {\bf51} (2019) 121.


\bibitem{Yang1}
Y. Yang, Electromagnetic asymmetry, relegation of curvature singularities of charged black holes, and cosmological equations of state in view of the Born--Infeld theory, 
{\em Class. Quant. Gravity} {\bf39} (2022) 195007.
arXiv: 2104.07051.


\bibitem{Yang2}
Y. Yang, Dyonically charged black holes arising in generalized Born--Infeld  theory of electromagnetism, {\em Ann. Phys.} {\bf443} (2022) 168996. arXiv: 2204.11313. 

\bibitem{Yang3}
Y. Yang, Dyonic matter equations, exact point-source solutions, and charged black holes in generalized Born--Infeld theory,  {\em Phys. Rev.} D {\bf107} (2023) 085007.  arXiv: 2208.09737.


\bibitem{Jana}
S. Jana and S. Kar,
Born--Infeld cosmology with scalar Born--Infeld matter,
{\em Phys. Rev.} D {\bf94} (2016) 064016.

\bibitem{Kam}
A. Kamenshchik, C. Kiefer, and N. Kwidzinski,
Classical and quantum cosmology of Born--Infeld type models, {\em Phys. Rev.} D (2016) 083519.

\bibitem{Nov}
M. Novello, M. Makler, L. S. Werneck, and C. A. Romero,
Extended Born--Infeld dynamics and cosmology,
{\em Phys. Rev.} D {\bf71} (2005) 043515.

\bibitem{JHOR}
J. B. Jimenez, L. Heisenberg, G. J. Olmo, and D. Rubiera-Garcia,
Born--Infeld inspired modifications of gravity, {\em Phys. Repts.} {\bf727} (2018) 1--129.

\bibitem{Yang4}
Y. Yang, Nonlinear problems inspired by the Born--Infeld theory of electrodynamics, {\em Adv. Nonlinear Stud.}, to appear. arXiv:2304.08236.

\bibitem{Kr1}
S. I. Kruglov,
On generalized Born--Infeld electrodynamics, {\em J. Phys.} A {\bf43} (2010) 375402.

\bibitem{Soleng}
H. H. Soleng,
Charged black points in General Relativity coupled to the logarithmic $U(1)$ gauge theory, {\em Phys. Rev.} D {\bf52}
(1995) 6178--6181.


\bibitem{Fe}
 J. A. Feigenbaum, P. G. O. Freund, and M. Pigli, Gravitational analogues of non-linear
Born electrodynamics, {\em Phys. Rev.} D {\bf57} (1998) 4738--4744.

\bibitem{AM}
P. N. Akmansoy and L. G. Medeiros, Constraining Born--Infeld-like nonlinear electrodynamics using hydrogen's ionization energy,
{\em Euro. Phys. J.} C {\bf78} (2018) 143.

\bibitem{Gaete}
P. Gaete, J. A. Helay\"{e}l-Neto, and L. P. R. Ospedal,
Coulomb's law modification driven by a logarithmic electrodynamics, {\em Europhys. Lett.} {\bf125}
(2019) 51001.

\bibitem{K6}
S. I. Kruglov,
Dyonic black holes with nonlinear logarithmic electrodynamics, {\em Grav. Cosmol.} {\bf25} (2019) 190--195.


\bibitem{H1}
S. H. Hendi, Asymptotic charged BTZ black hole solutions, {\em J. High. Energy Phys.} {\bf 03} (2012)
065.

\bibitem{H2}
 S. H. Hendi, Asymptotic Reissner--Nordstr\"{o}m black holes, {\em Ann. Phys.} {\bf333} (2013) 282--289.

\bibitem{Kr2}
S. I. Kruglov,
Notes on Born--Infeld-type electrodynamics, {\em Mod. Phys. Lett.} A {\bf32} (2017) 1750201.

\bibitem{CG}
R. M. Corless,  G. H. Gonnet, D. E. G. Hare, D. J. Jeffrey, and D. E. Knuth,  On the Lambert $W$ function, {\em
Adv. Comput. Math.} {\bf 5} (1996) 329--359.


\bibitem{Chow}
T. Y. Chow,
What is a closed-form number? {\em Amer. Math. Monthly} {\bf106} (1999) 440--448.

\bibitem{De}
V. A. De Lorenci, R. Klippert, M. Novello, and J. M. Salim,
Nonlinear electrodynamics and FRW cosmology, {\em Phys. Rev.} D {\bf65} (2002) 063501.

\bibitem{K2007}
S. I. Kruglov,
Vacuum birefringence from the effective Lagrangian of the electromagnetic field,
{\em Phys. Rev.} D {\bf75} (2007) 117301.

\bibitem{GS}
 R. Garcia-Salcedo, T.  Gonzalez, and I. Quiros,
No compelling cosmological models come out of magnetic universes which are based in nonlinear electrodynamics,
{\em Phys. Rev.} D {\bf89} (2014) 084047.

\bibitem{C2015}
C. V. Costa, D. M. Gitman,  and A. E. Shabad,
Finite field-energy of a point charge in QED,
{\em Phys. Scripta} {\bf90} (2015) 074012.

\bibitem{K2017}
S. I. Kruglov,
Remarks on Heisenberg--Euler-type electrodynamics,
{\em Mod. Phys. Lett.} A {\bf32} (2017) 1750092.

\bibitem{Lamb}
H. Lamb, {\em Hydrodynamics}, 6th edition, Cambridge Univ. Press, Cambridge, U. K., 1932.

\bibitem{Evans}
L. C. Evans, {\em Partial Differential Equations}, Amer. Math. Soc., Providence, Rhode Island, 2010.

\bibitem{GT}
D. Gilbarg and N. S. Trudinger, {\em Elliptic Partial Differential Equations of Second Order}, Springer-Verlag, Berlin and New York, 2001.

\bibitem{SY}
J. Spruck and Y. Yang,  Charged cosmological dust solutions of the coupled Einstein and Maxwell equations, {\em Discrete Cont. Dynam. Sys.} {\bf28} (2010) 567--589.




\end{thebibliography}
\end{document}